\documentclass{aa}
\usepackage[dvips]{graphicx}
\usepackage{natbib}
\usepackage{txfonts}
\usepackage{color}

\newcommand{\stimes}{\negthinspace\times\negthinspace}
\begin{document}
   \title{Radiative hydrodynamics simulations of red supergiant stars.
III. Spectro-photocentric variability, photometric variability, and
consequences on Gaia measurements}
\titlerunning{Spectro-photocentric and photometric variability of red supergiant stars}

   \author{A. Chiavassa
          \inst{1,2}
          \and
          E. Pasquato \inst{1} 
          \and
	 A. Jorissen\inst{1}
	 \and
	 S. Sacuto \inst{3}
\and
 C. Babusiaux\inst{4}
          \and
          B. Freytag\inst{5,6,7}
          \and
          H.-G. Ludwig\inst{8}
          \and
          P. Cruzal\`ebes \inst{9}
         \and
         Y. Rabbia \inst{9}
         \and
         A. Spang \inst{9}
         \and
         O. Chesneau \inst{9}         
          }
   \offprints{A. Chiavassa}

   \institute{Institut d'Astronomie et d'Astrophysique, Universit\'e Libre de Bruxelles, CP. 226, Boulevard du Triomphe, B-1050 Bruxelles, Belgium\\
              \email{achiavas@ulb.ac.be}
              \and
              Max-Planck-Institut f\"{u}r Astrophysik, Karl-Schwarzschild-Str. 1, Postfach 1317, D-85741 Garching b. M\"{u}nchen, Germany
          \and
         Department of Astronomy, University of Vienna,
         T\"urkenschanzstrasse 17, A-1180 Wien, Austria  
          \and
          GEPI, Observatoire de Paris, CNRS, Universit\'e Paris Diderot, Place Jules Janssen F-92190 Meudon, France
         \and
         Universit\'e de Lyon, F-69003 Lyon, France; Ecole Normale Sup\'erieure de Lyon, 46 all\'ee d'Italie, F-69007 Lyon, France; CNRS, UMR 5574, Centre de Recherche Astrophysique de Lyon; Universit\'e Lyon 1, F-69622 Villeurbanne, France
         \and
         Department of Physics and Astronomy,
         Division of Astronomy and Space Physics,
         Uppsala University,
         Box 515, S-751~ 20 Uppsala,
         Sweden
          \and
          Istituto Nazionale di Astrofisica, Osservatorio Astronomico
          di Capodimonte, Via Moiariello 16, I-80131 Naples, Italy
          \and
         Zentrum f\"ur Astronomie der Universit\"at Heidelberg, Landessternwarte, K\"onigstuhl 12, D-69117 Heidelberg, Germany
         \and
         UMR 6525 H. Fizeau, Univ. Nice Sophia Antipolis, CNRS, Observatoire de  la C\^{o}te d'Azur, Av. Copernic, F-06130 Grasse, France
             }

   \date{Received; accepted }

  \abstract
   {It has been shown that convection in red supergiant stars (RSG) gives rise to
large granules causing surface inhomogeneities together with shock waves in the photosphere. The resulting  motion of the photocenter (on time scales ranging from months to years) could possibly have adverse effects on the parallax determination with Gaia.}
   {We explore the impact of the granulation on the photocentric and
     photometric variability. We quantify these effects in order to better characterize the error possibly altering the parallax.}
   {We use 3D radiative-hydrodynamics (RHD) simulations of convection with
 CO5BOLD and the
   post-processing radiative transfer code OPTIM3D to compute intensity
maps and spectra in the Gaia $G$ band [325 -- 1030~nm]. 
}
   {We provide astrometric  and photometric predictions from 3D
     simulations of RSGs that are used to evaluate the possible degradation of the astrometric parameters of evolved stars
    derived by  Gaia. We show in particular from RHD simulations that a supergiant like Betelgeuse exhibits a photocentric noise characterised by a standard deviation of the order of 0.1~AU. The number of bright giant and supergiant stars whose Gaia parallaxes will be altered by the photocentric noise  ranges from a few tens to several thousandths, 
depending on the poorly known relation between the size
    of the convective cells  and the atmospheric pressure scale height
    of supergiants, and to a lower extent, on the adopted prescription for
    galactic extinction. In the worst situation,  the degradation of the astrometric fit due to the presence of this photocentric noise 
will be noticeable up to about 5~kpc for the brightest supergiants. Moreover, parallaxes of Betelgeuse-like supergiants are affected by a error of the order of a few percents.
 We also show that the photocentric
     noise, as predicted by the 3D simulation, does account for a
     substantial part of the supplementary 'cosmic noise' that affects
      Hipparcos measurements of Betelgeuse and Antares. 
   }
   {}

               
\maketitle
               
%

\section{Introduction}

The main goal of the Gaia mission \citep{2001A&A...369..339P,2008IAUS..248..217L} is to
determine high-precision astrometric parameters (i.e., positions,
parallaxes, and proper motions) for one billion objects with apparent magnitudes in the range $5.6 \le V \le 20$. These data along
with multi-band and multi-epoch photometric
and spectrocopic data will allow to reconstruct the formation history,
structure, and evolution of the Galaxy. Among all the objects that will
be observed, late-type stars present granulation-related variability
that is considered, in this context, as "noise" that must be
quantified in order to better characterize any resulting error on the parallax determination. A previous work by \cite{2006A&A...445..661L} has shown that
effects due to the granulation in red giant stars are not likely to be
important except for the
extreme giants.\\
Red supergiant (RSG) stars are late-type stars with masses between 10 and 40 M$_{\odot}$. They have effective temperature $T_{\rm eff}$ ranging from 3450 (M5) to 4100 K (K1), luminosities in the range 2000 to 300\,000 L$_{\odot}$, and radii up to 1500 R$_{\odot}$ \citep{2005ApJ...628..973L}. Their luminosities place them among the brightest
stars, visible up to very large distances. Based on detailed
radiation-hydrodynamics (RHD) simulations of RSGs
(\citealp{2002AN....323..213F} and
\citealp{Freytag2008A&A...483..571F}), \cite{2009A&A...506.1351C}
(Paper~I hereafter) and \cite{2010A&A...515A..12C} (Paper~II hereafter) show that these stars are characterized by vigorous
convection which imprints a pronounced granulation pattern on the
stellar surface. In particular, RSGs give rise to large granules
comparable to the stellar radius in the $H$ and $K$ bands, and an irregular pattern in the optical region. 

This paper is the third in the series aimed at exploring the
convection in RSGs. The main purpose is to extract photocentric and
photometric predictions that will be used to estimate the number of
RSGs, detectable by Gaia, for which the parallax measurement will be
affected by the displacements of their photometric centroid (hereafter "photocenter").

\section{RHD simulations of red supergiant stars}

The numerical simulation used in this work has been computed
using CO$^5$BOLD \citep{2002AN....323..213F,
  Freytag2008A&A...483..571F}. The model, deeply analyzed in Paper~I,
has a mass of 12 $M_{\odot}$, employs an equidistant numerical mesh
with 235$^3$ grid points with a resolution of
8.6~$R_{\odot}$ (or 0.040 AU), a luminosity average over spherical shells and over
time (i.e., over 5 years) of $L=93\,000\pm1300$~$L_{\odot}$, an effective temperature of
$T_{\rm{eff}}=3490\pm$13~K, a radius of $R=832\pm0.7$~$R_{\odot}$,
and a surface gravity $\log g=-0.337\pm0.001$. The uncertainties are
measures of the temporal fluctuations. This is our most successful
RHD simulation so far because it has stellar parameters closest to
Betelgeuse ($T_{\rm{eff}}=3650$~K, $\log g=0.0$, \citeauthor{2005ApJ...628..973L},
\citeyear{2005ApJ...628..973L},  or $\log g = -0.3$, \citeauthor{2008AJ....135.1430H}, \citeyear{2008AJ....135.1430H}). We stress that 
the surface gravity of Betelgeuse is poorly known, and this is not without consequences for the analysis that will be presented in Sect.~\ref{Sect:Gaia-frequency} (see especially Fig.~\ref{fig_gaia2bis}).

For the computation of the intensity maps and spectra based on
snapshots from the RHD simulations, we used the code
OPTIM3D (see Paper~I) which takes into account the Doppler shifts caused by the
convective motions. The radiative transfer is computed in detail using pre-tabulated extinction coefficients per unit mass generated with MARCS (\citealp{2008A&A...486..951G}) as a function of temperature, density and wavelength for the solar composition \citep{2006CoAst.147...76A}. The tables include the
same extensive atomic and molecular data as the MARCS models. They
were constructed with no micro-turbulence broadening and the
temperature and density distributions are optimized to cover the values
encountered in the outer layers of the RHD
simulations. 

\section{Predictions}

In this Section we provide a list of predictions from 3D simulations that are related to the Gaia astrometric and photometric measurements.

\subsection{Photocenter variability}\label{sect:photocenter}

\begin{figure}
   \centering
         \includegraphics[width=0.9\hsize]{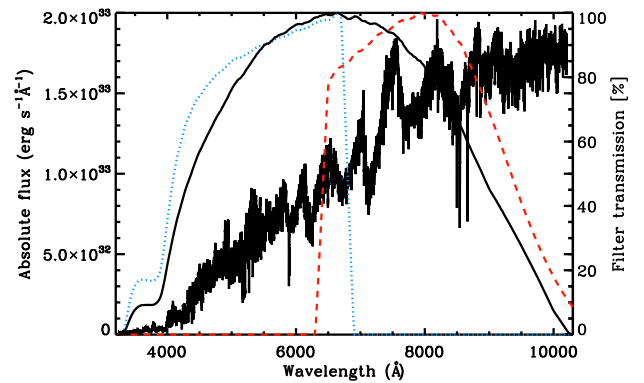} \\
      \caption{The transmission curve of the Gaia $G$ band white light
        passband (solid black line), the blue (dotted blue line) and
        red (dashed red line) photometric filters \citep{2010A&A...523A..48J,2007ASPC..364..215J} together with the synthetic spectrum computed from the RHD simulation described in the text.
           }
        \label{fig1}
   \end{figure}

\begin{figure*}
   \centering
   \includegraphics[width=1\hsize]{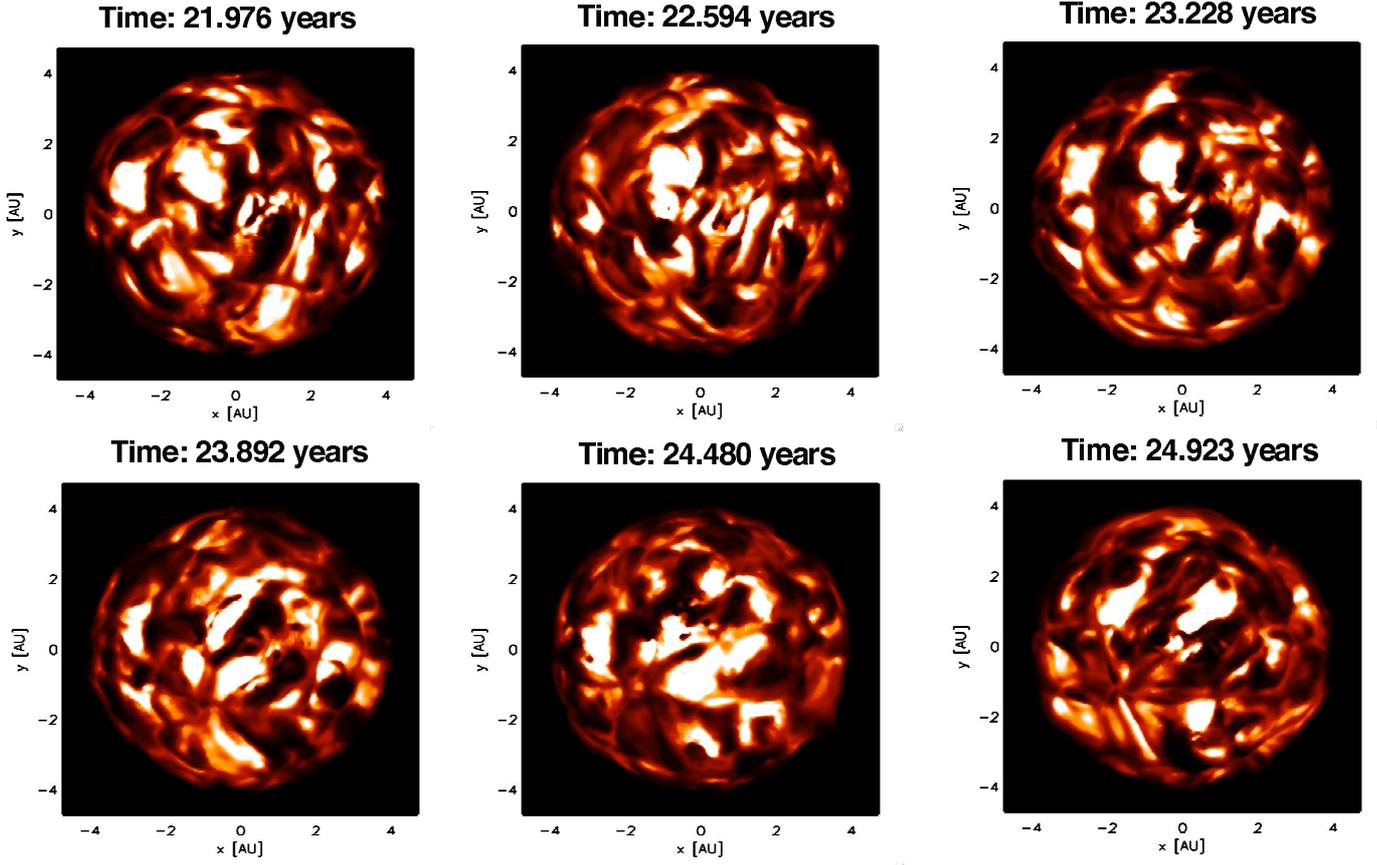}
      \caption{Maps of the linear intensity (the range is [0 -- 230000]
        erg/s/cm$^2$/\AA ) in the Gaia $G$ band. Each panel corresponds
        to a different snapshot of the model described in the text
        with a step of about 230 days ($\approx5$ years covered by
        the simulation). }
         \label{fig2}
   \end{figure*}

We computed spectra and intensity maps in the Gaia $G$ band
 for the whole
simulation time sequence, namely $\sim\!\!5$\, years with snapshots 
$\approx23$ days apart. 
The corresponding spectrum is presented in Fig.~\ref{fig1} and the images in Fig.~\ref{fig2}. 

Paper~II showed that the intensity maps in the optical
region show high-contrast patterns characterized by dark spots and bright
areas. The
brightest areas exhibit an intensity
  50 times brighter than the dark
ones with strong changes over some weeks. Paper~II reported robust interferometric comparisons of hydrodynamical simulations with existing observations in
the optical and $H$ band regions, arguing for the presence of convective cells of various sizes on the red
supergiant Betelgeuse. The Gaia $G$ band images (Fig.~\ref{fig2}) are
comparable to what has been found in Paper~II. The resulting surface pattern, though related to
the underlying granulation pattern, is also connected to dynamical effects. In
fact, the emerging intensity depends on (i) the opacity run through
the atmosphere (and in red supergiants, TiO molecules produce strong
absorption at these
wavelengths; see spectrum in Fig.~\ref{fig1}) and on (ii) the shocks and waves which
dominate at optical depths smaller than~1. 

\begin{figure}
   \includegraphics[width=0.99\hsize]{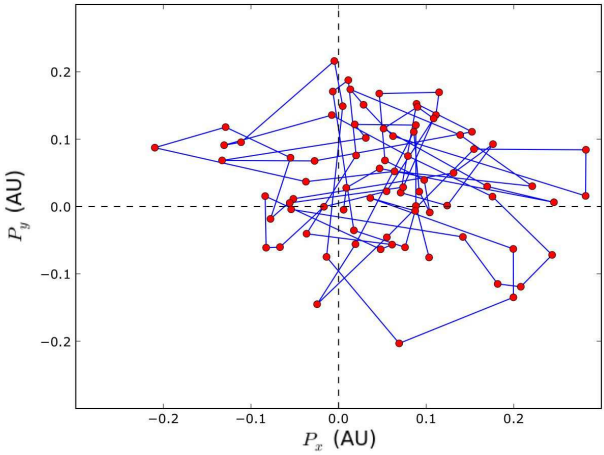}
      \caption{
Photocenter position, computed from
        the snapshots of Fig.~\ref{fig2}, in the Gaia-$G$
        band filter. The total simulated time is $\approx$5 years
        and the snapshots are 23 days apart. The snapshots are
        connected by the line segments. The dashed lines intersect at the
      position of the geometrical center of the images. Note that the photocentric shift stays in the first quadrant for most of the 5~yr simulation, and reflects the long lifetime of the large  convective cell best visible in the infrared $H$ band (Fig.~\ref{figHband}).}
         \label{fig3}
   \end{figure}

The surface appearance of RSGs in the Gaia $G$ band affects strongly the position of the photocenter and cause temporal
fluctuations. The position of the photocenter is given as the intensity-weighted mean of the $x-y$ positions of all emitting points tiling the visible stellar surface according to:

\begin{eqnarray}
P_x=\frac{\sum_{i=1}^{N} \sum_{j=1}^{N} I(i,j)*x(i,j)}{\sum_{i=1}^{N} \sum_{j=1}^{N} I(i,j)} \\
P_y=\frac{\sum_{i=1}^{N} \sum_{j=1}^{N} I(i,j)*y(i,j)}{\sum_{i=1}^{N} \sum_{j=1}^{N} I(i,j)}
\end{eqnarray}

where $I\left(i,j\right)$ is the emerging intensity for the grid
point $(i,j)$ with coordinates $x(i,j)$, $y(i,j)$ of the
simulation, and $N=235$ is the total number of grid points.
Fig.~\ref{fig3} shows that the photocenter excursion is
large, since it goes from
0.005 to 0.3 AU over 5 years of simulation (the stellar
radius is $\approx$ 4 AU, Fig.~\ref{fig2}). The temporal average value
of the photocenter displacement is $\langle P\rangle=\langle (P_x^2 + P_y^2)^{(1/2)}\rangle
=0.132$~AU, and $\sigma_P =  0.065$~AU. 

\begin{figure}
\centering
   \includegraphics[width=0.8\hsize]{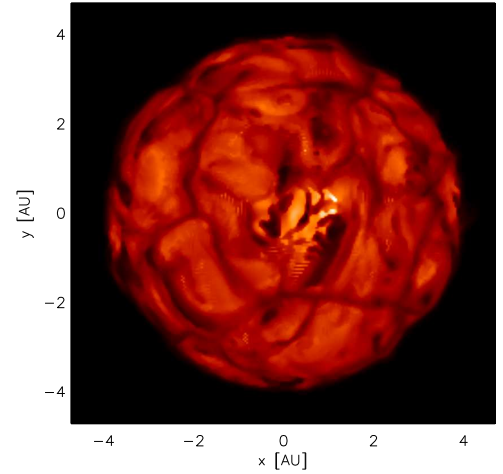}
      \caption{Map of the linear intensity in the
  IONIC filter ($H$ band as described in Paper~I). The range is
  [0 - $3.1\times10^5$]\,erg\,cm$^{-2}$\,s$^{-1}$\,{\AA}$^{-1}$. The snapshot corresponds to the top left snapshot at $t=21.976$~yr in Fig.~\ref{fig2}. The large convective cell visible in this Figure is swamped in smaller-scale photospheric structures in the Gaia $G$ band images.  
}
         \label{figHband}
   \end{figure}

At this point, it is important to define the characteristic time scale of the convective-related surface structures. RHD simulations show that RSGs are characterized by two characteristic time scales:
\begin{itemize}
\item[(i)]  the surface of the RSG is covered by a few large convective cells with a size of about
1.8--2.3~AU  ($\approx$60$\%$ of the stellar radius) that evolve on a {\it time-scale of years} (see Fig.~\ref{figHband} and Paper~I). This is visible in the infrared, and particularly in the $H$ band where the H$^{-}$ continuous opacity minimum
occurs and consequently, the continuum-forming region is more evident.
\item[(ii)] In the optical region, as in Fig.~\ref{fig2}, short-lived ({\it a few weeks to a few months})
small-scale (about 0.2--0.5~AU, $\approx$10$\%$ of the stellar radius) structures appear. They result from the opacity run and dynamics at optical depths smaller than~1 (i.e., further up in the atmosphere with respect to the continuum-forming region). 
\end{itemize}

Both time scales have an effect on the photocenter excursion during the 5 years covered by the simulation.
On one hand, the value of $\sigma_P$ is mostly fixed by the short time scales corresponding to the small atmospheric structures. On the other hand, the fact that  $\langle P_x\rangle$ and $\langle P_y\rangle$ do not average to 
zero (according to Fig.~\ref{fig3}, the photocenter stays most of the time in the same quadrant, due to the presence of the large convective cell which is visible in the $H$ band; see Fig.~\ref{figHband} and Paper~I) indicates that the 5 years period covered by the simulation is  not yet long enough with respect to the characteristic time scale  
of the large-scale (continuum) cells. 

\begin{figure}
   \includegraphics[width=1.0\hsize]{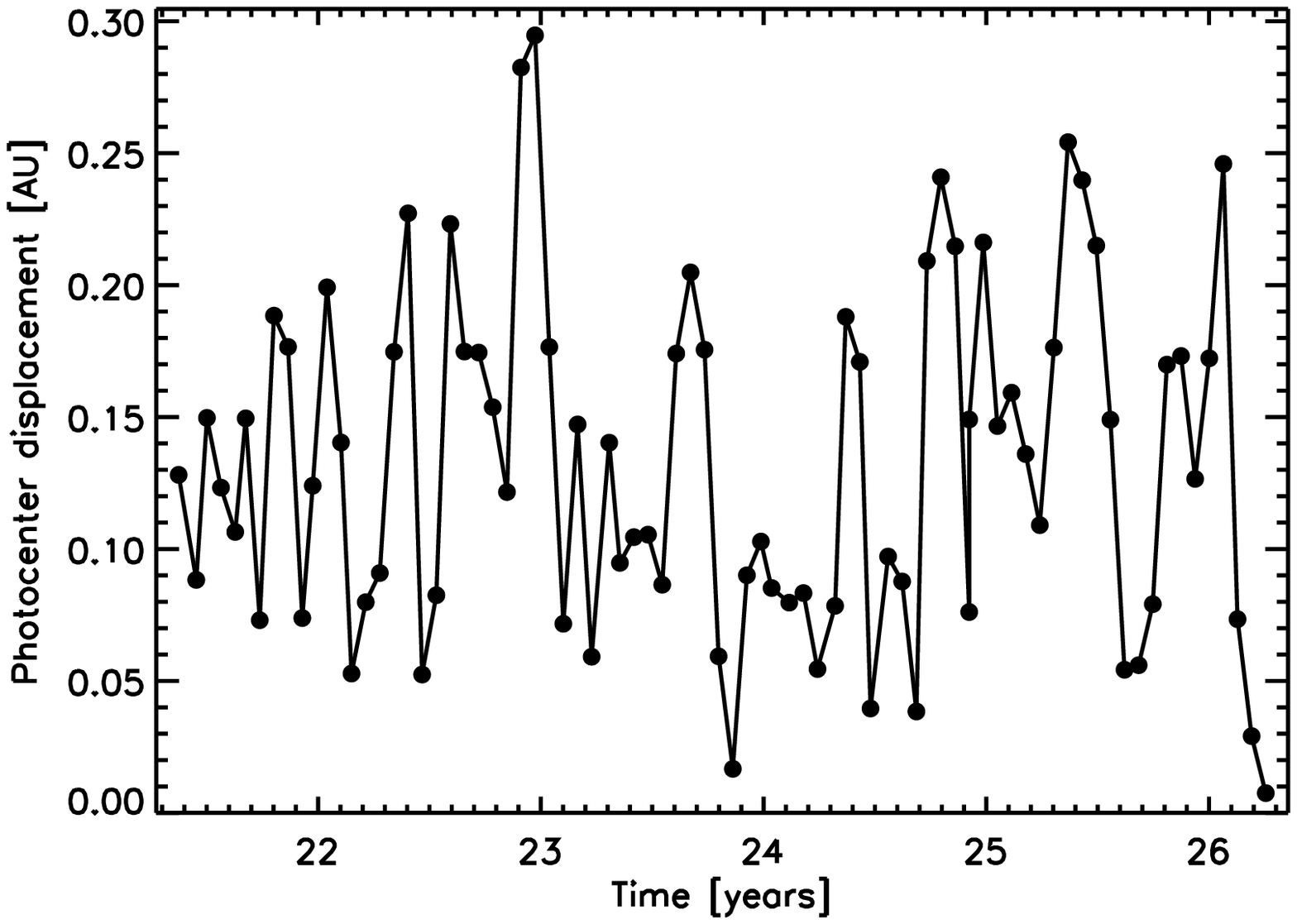}
   \includegraphics[width=1.0\hsize]{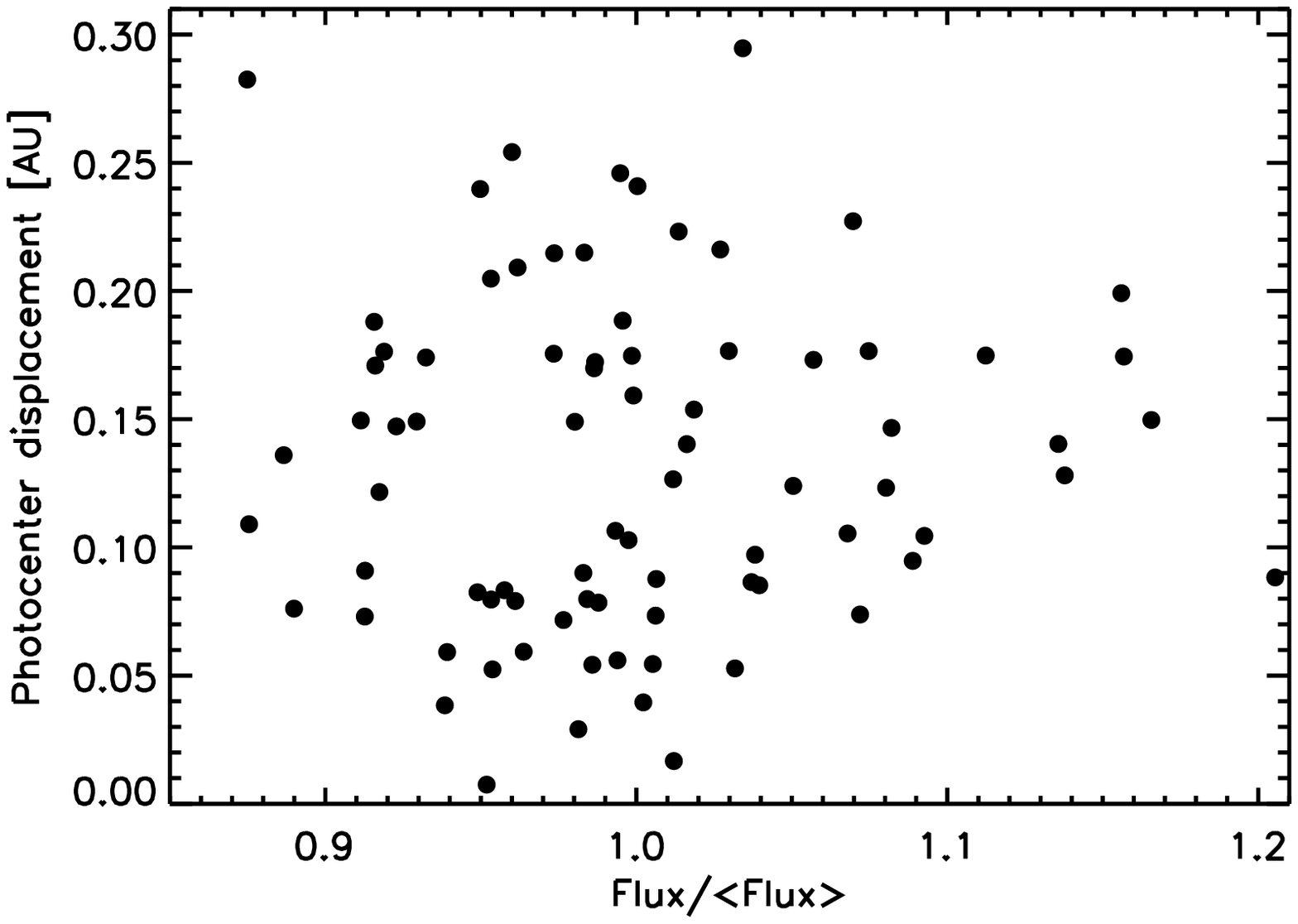}
      \caption{\emph{Top panel:} modulus of the photocenter
        displacement as a function of time. \emph{Bottom panel:}
        photocenter displacement as a function of the integrated flux
        in the Gaia $G$ band, $\int_{3250\AA}^{10300\AA}F_\lambda\; \mathrm{d}\lambda$, normalized by the temporal average integrated flux.}
         \label{fig4}
   \end{figure}

The top panel of Fig.~\ref{fig4} shows the temporal photocenter
displacement over the $\approx5$~years of simulation, which is
comparable to the total length of the Gaia mission. As seen in the Figure,
for $t<22$~yr, the random displacement is small and increases to a maximum
value of 0.30~AU at $t \sim 23$~yr. 

In relation with the astrometric implications of this photocentre displacement, which will be discussed in Sect.~\ref{Sect:Photocentervariability}, it must be stressed that neither  $\langle P\rangle$ nor $\sigma_P$ (the latter corresponding to the time sampling of the photocentric motion with a rather arbitrary time interval of 23~days) are the relevant quantities; it is instead the standard deviation of $P$ sampled as Gaia will do (both timewise and directionwise) which turns out to be relevant. This quantity is computed below.

The bottom panel of Fig.~\ref{fig4} shows that there is no obvious correlation between the photocenter
variability and the emerging intensity integrated in the Gaia $G$
band. \cite{2006A&A...445..661L} showed analytically that this lack of
correlation is to be expected.

\begin{figure}
   \includegraphics[width=1.\hsize]{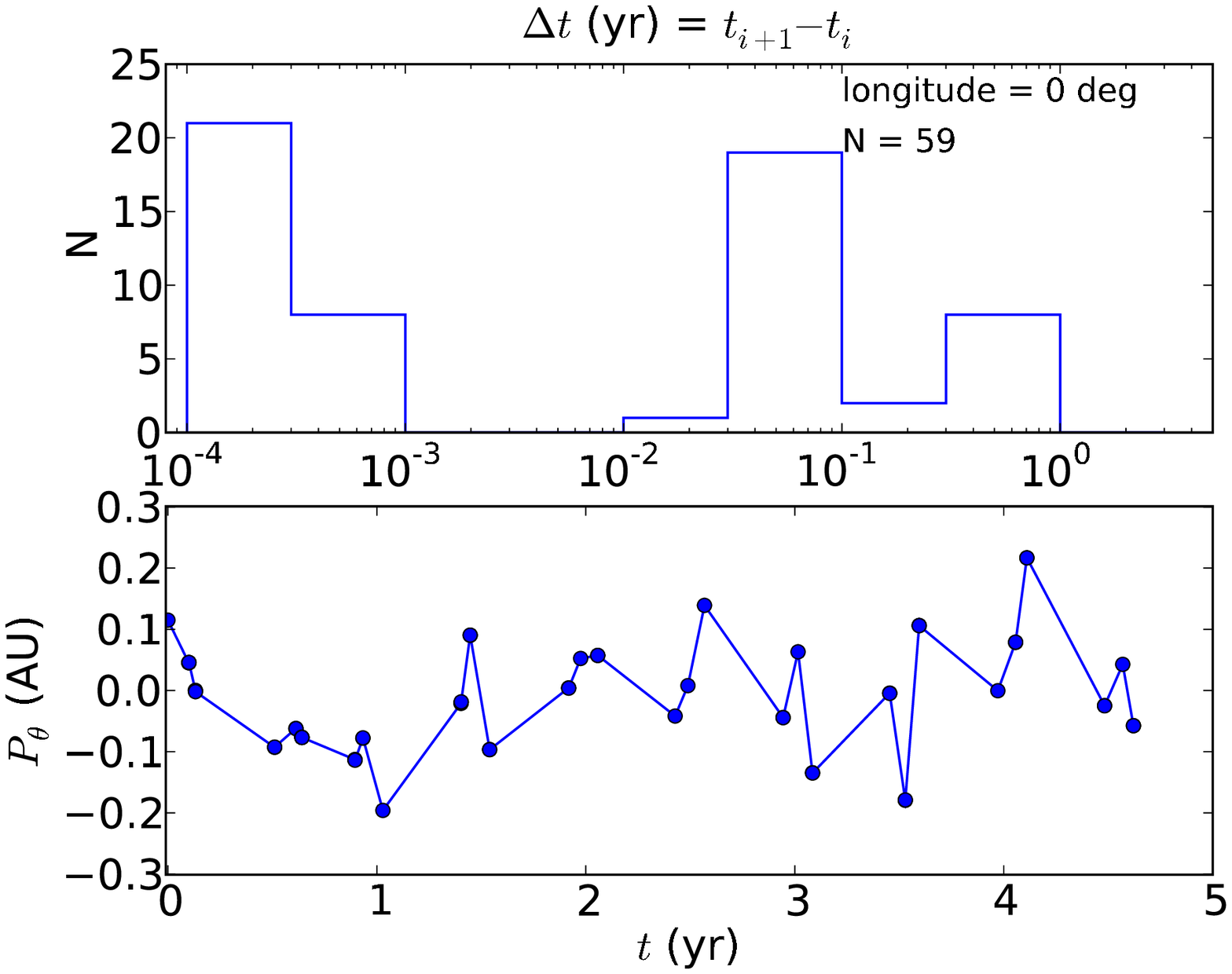}
   \includegraphics[width=1.\hsize]{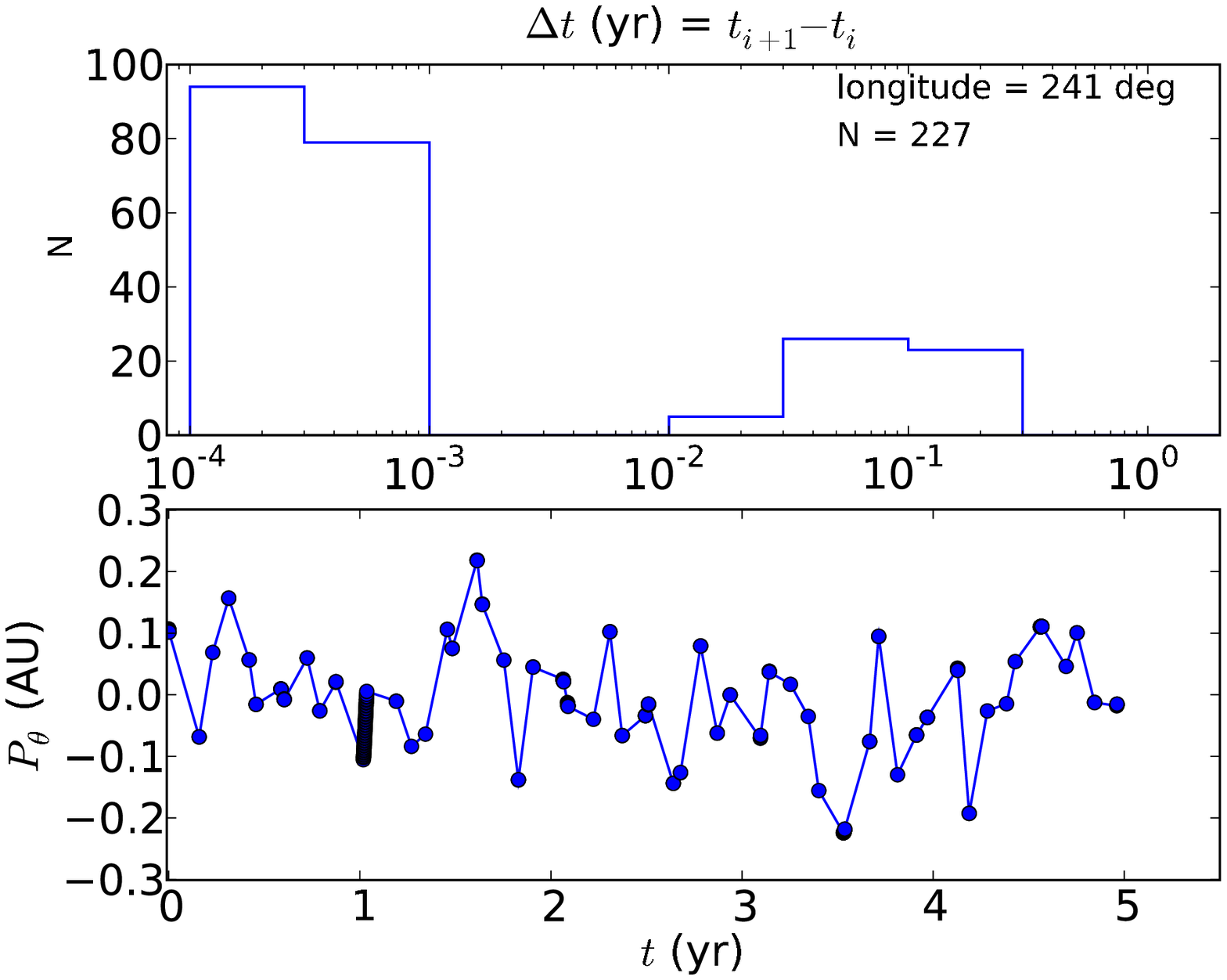}
      \caption{
      The along-scan photocenter displacement $P_{\theta}$ (in AU) against time for two
      different samplings of the photocenter displacement of Fig.~\ref{fig3}, corresponding to the Gaia 
      scanning law applied to stars  located along the galactic plane at longitudes of $0^\circ$ and $241^\circ$, 
as indicated on the figures.  
     The top panel of each pair provides  the distribution of time intervals between successive measurements.
  }
         \label{fig_PAL}
   \end{figure}

Gaia will scan the sky, observing each object on average 70-80
times. The main information that will be used to determine the
astrometric characteristics of each stars will be the along-scan (AL)
measurement. This is basically the projection of the star position
along the scanning direction of the satellite with respect to a known
reference point. By fitting those data through a least square
minimization, the position, parallax and proper motion of the star can be
derived. The possibility of extracting these parameters is ensured by Gaia's complex
scanning law\footnote{See http://www.rssd.esa.int/index.php?project=GAIA\&page=picture \_of\_the\_week\&pow=13} 
which guarantees that every star is observed from many different
scanning angles.\\
\indent In presence of surface brightness asymmetries the photocenter position will no more coincide with the barycenter of the star and its position will change as the surface pattern changes with time. The result of this phenomenon is that the AL measurements of Gaia will reflect proper motion, parallactic motion (that are modeled to obtain the astrometric parameters of the star) and photocentric motion of convective origin.
The presence of the latter will be regarded as a source of additional noise.\\
\indent The impact of those photocenter fluctuations on the astrometric
quantities will depend on several parameters, some of which are the stellar
distance and the time sampling (fixed by the scanning law) of the
photocentric motion displayed on Fig.~\ref{fig3}. To better assess this impact,
we proceeded as follows.
The Gaia Simulator \citep{2005ESASP.576..357L} was used to derive scanning angles and time sampling for stars regularly spaced (one degree apart) along the galactic plane where the supergiants are found.
We computed the
photocenter coordinates at the Gaia transit times, by linear interpolation of the photocenter
    positions of the model (as provided by
Fig.~\ref{fig3}), after subtracting $\langle P_x \rangle$ (=0.055~AU) and  $\langle P_y \rangle$ (=0.037~AU; as we will explain below, a constant photocentric offset has no astrometric impact on the parallax). 
We then computed
their projection on the AL direction, which we denote
$P_{\theta}$, $\theta$ being the position angle along the scanning
direction on the sky. This projection $P_{\theta}$ relates to the modulus $P$ of the photocenter vector plotted in
Fig.~\ref{fig4} through the relation
\begin{equation}
\label{Eq:Ptheta}
P_\theta = P \cos(\theta - \theta_P) \;\;\;\mathrm{with}\;\;\; \tan \theta_P = \left(P_y - \langle P_y \rangle\right) / \left(P_x -  \langle P_x \rangle \right),
\end{equation}
and similarly, we define
\begin{equation}
\label{Eq:Ptheta'}
P'_\theta = P \cos(\theta - \theta_P')\;\;\; \mathrm{with} \;\;\;\tan \theta_P' = P_y  / P_x .
\end{equation}
The resulting run of the standard deviation of the photocenter
displacement with time for two representative stars (one located at $l = 0^\circ$ with  59 transits and the other located at $l = 241^\circ$ with 227 transits) is shown on Fig.~\ref{fig_PAL}, which reveals that the time
sampling is, as expected, strongly dependent upon the star position on
the sky. The transits separated by $2-6\stimes 10^{-4}$\,yrs correspond to the star being observed in succession by the two fields of view (separated by 106.5 degrees 
on the sky) by the satellite spinning at a rate of 6 hours per cycle, whereas the longer intervals are fixed by the satellite precession rate.

\begin{figure}
 \includegraphics[width=1.0\hsize]{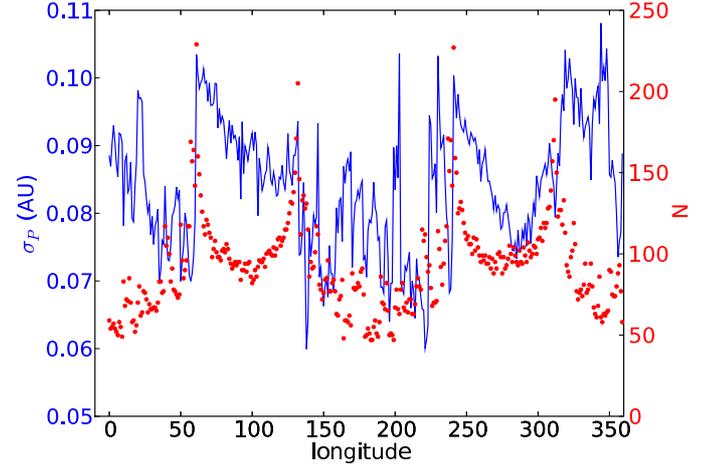}
\caption{\label{Fig:sigmaP}
The run of $\sigma_{P_\theta}$ (expressed in AU; solid blue curve and left-hand scale) with galactic longitude for stars located along the galactic plane, having a 
number of transit observations given by the red dots (and right-hand scale). 
}
\end{figure}

Finally, we computed the standard deviation of those projections, and obtained
$\sigma_{P_\theta}$ values ranging from 0.06 to 0.10~AU (Fig~\ref{Fig:sigmaP}), with $\langle |P_{\theta}| \rangle$ ranging from $1\stimes 10^{-4}$ to $8\stimes 10^{-2}$\,AU.
In the remainder of this paper, we will adopt $\sigma_{P_\theta}=0.08$\,AU. 
This quantity, which represents about 2.0\% of the stellar radius ($\approx$4~AU; Sect.~2),
is a measure of the mean photocenter
noise induced by the convective cells in the model, and it is this
value which needs to be compared with the Gaia or Hipparcos
measurement uncertainty
to evaluate the impact of granulation noise on the astrometric
parameters. This will be done in
Sections~\ref{Sect:Photocentervariability} and \ref{Sect:Hipparcos}.

We note that $\sigma_{P_\theta} = 0.08$~AU is in fact larger than $\sigma_P = 0.065$~AU, and this can be understood as follows. 
First, from Eq.~(\ref{Eq:Ptheta'}) and basic  statistical principles, the following relation may be easily demonstrated:
\begin{equation}
\sigma_{P'_\theta}^2 = 0.5 \;(\sigma_P^2 + \langle P \rangle^2),
\end{equation}
under the obvious hypothesis of statistical independence between the scanning directions $\theta$ and the photocentric positions $P_x, P_y$.
With $\langle P\rangle
=0.132$~AU, and $\sigma_P =  0.065$~AU obtained in Sect.~\ref{sect:photocenter}, the above relation predicts
$\sigma_{P'_\theta} = 0.10$~AU, in agreement with the actual predictions based on Eq.~(\ref{Eq:Ptheta'}). If one considers instead $P_\theta$ from Eq.~(\ref{Eq:Ptheta}) (thus projecting the 're-centered' photocentric displacement), there is a small reduction of the standard deviation 
according to
\begin{equation}
\sigma_{P_\theta}^2 = \sigma_{P'_\theta}^2 - 0.5 (\langle P_x \rangle^2 + \langle P_y \rangle^2),
\end{equation}
yielding $\sigma_{P_\theta} = 0.088$~AU, in agreement with the detailed calculations shown on Fig.~\ref{Fig:sigmaP}.


\subsection{Photometric variability}

\begin{figure}
  \begin{tabular}{c}
   \includegraphics[width=0.93\hsize]{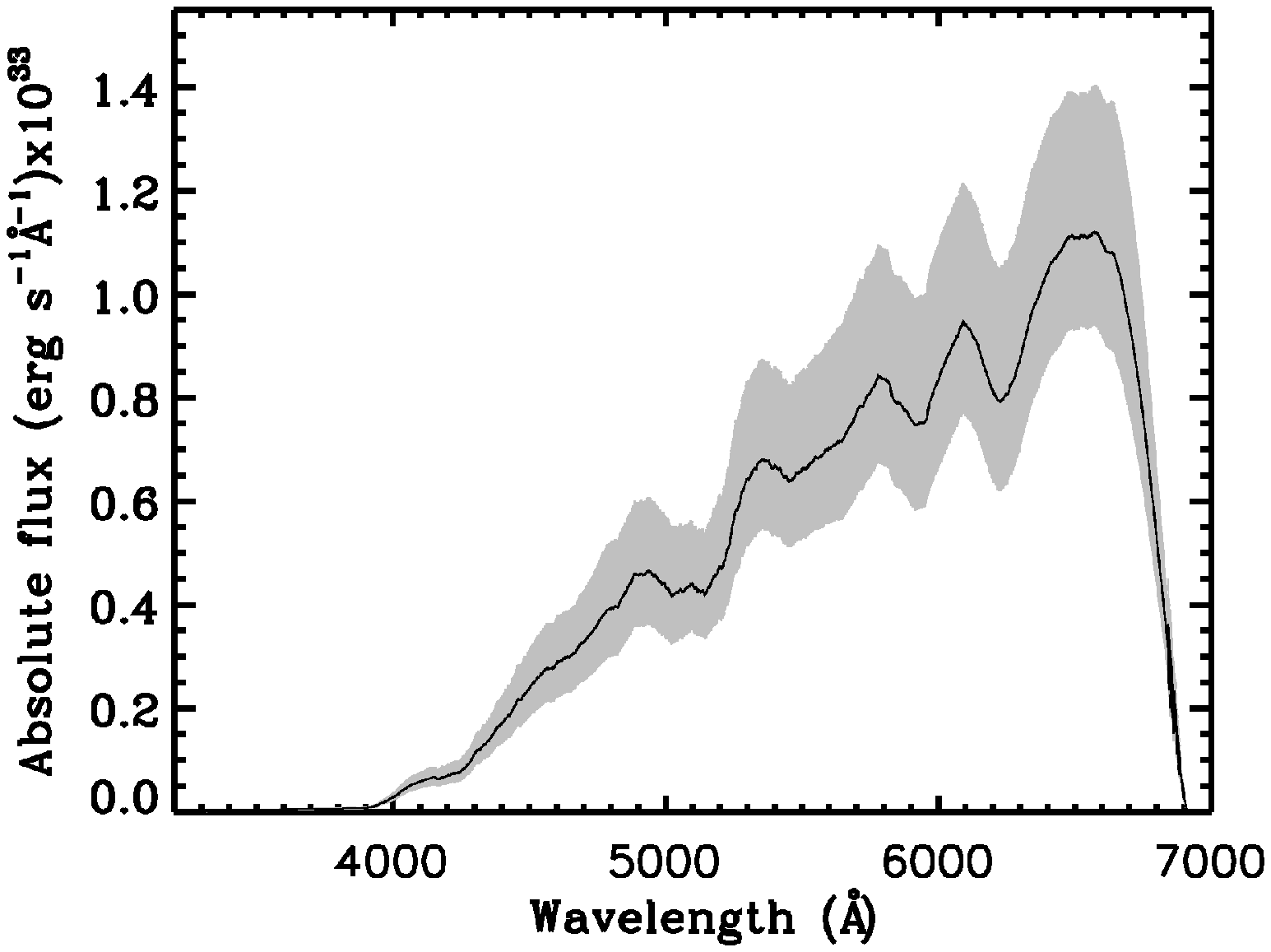}\\
   \includegraphics[width=0.9\hsize]{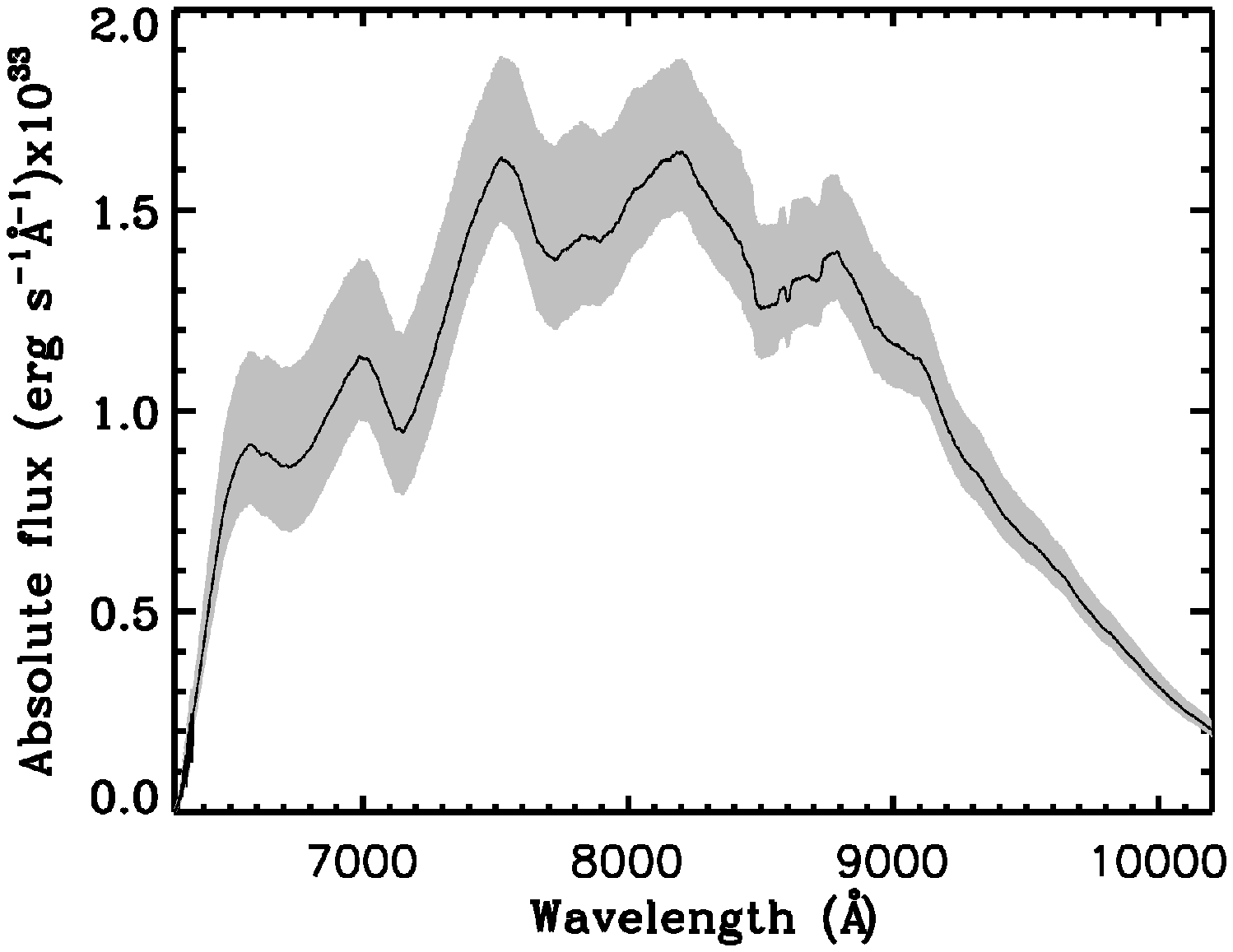}
\end{tabular}
      \caption{Spectral fluctuations in the blue and red Gaia
        photometric bands (Fig.~\ref{fig1}) for RSGs: the black curve is the average flux over $\approx$5 years covered by the
simulation, while the grey shade denotes the maximum and minimum fluctuations. The spectra have been smoothed to the Gaia spectral resolution \citep[$R\approx50$, ][]{2008PhST..133a4010T}.
  }
         \label{fig_foto1}
   \end{figure}

Another aspect of RSG variability can affect Gaia spectrophotometry. The blue and red photometric bands of Fig.~\ref{fig1}
produce two spectra of the observed source at low spectral resolution
($R\approx50$). The photometric system has the advantage of covering
continuously a wide range of wavelengths providing a multitude of
photometric bands, but it has the great disadvantage of being
extremely hard to calibrate in flux and wavelength. The photometric
system of Gaia will be used to characterize the star's effective temperature, surface gravity and metallicity \citep{2008PhST..133a4010T}. The vigorous convective motions and the resulting surface asymmetries of RSGs cause strong fluctuations in the spectra that will affect Gaia spectrophotometric measurements (Fig.~\ref{fig_foto1}). In the blue photometric range (top panel), the fluctuations go up to 0.28 mag and up to 0.15 mag in the red filter (bottom panel) over the
5 years of simulation. These values are of the same order as the standard
deviation of the visible magnitude excursion in the last 70 years for Betelgeuse,
0.28 mag (according to AAVSO\footnote{American Association of Variable Star Observers, www.aavso.org}). 

\begin{figure}
   \includegraphics[width=0.9\hsize]{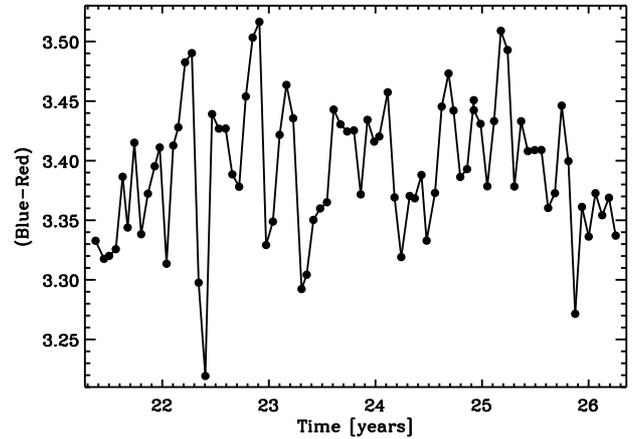}
      \caption{(blue - red) Gaia color index, computed in the blue and red photometric filters of Fig.~\ref{fig1}, as a function of time.
      }
         \label{GaiaColors}
   \end{figure}

The light curve of the simulation in the (blue - red) Gaia color index is displayed in Fig.~\ref{GaiaColors}. The temporal average value of the color index is (blue -red)$=3.39\pm0.06$ at one sigma and there are some extreme values at, for example, $t\approx22.4$~yr, $t\approx23$~yr, and $t\approx25.2$~yr. 

Therefore, the uncertainties on [Fe/H], $T_{\rm{eff}}$, and $\log g$ given by \cite{2010MNRAS.403...96B}
for stars with $G<15$ should be revised upwards for RSGs due to temporal fluctuations from convection.

\subsection{Direct imaging and interferometric observables}

\begin{figure*}
\centering
  \begin{tabular}{cc}
   \includegraphics[width=0.35\hsize]{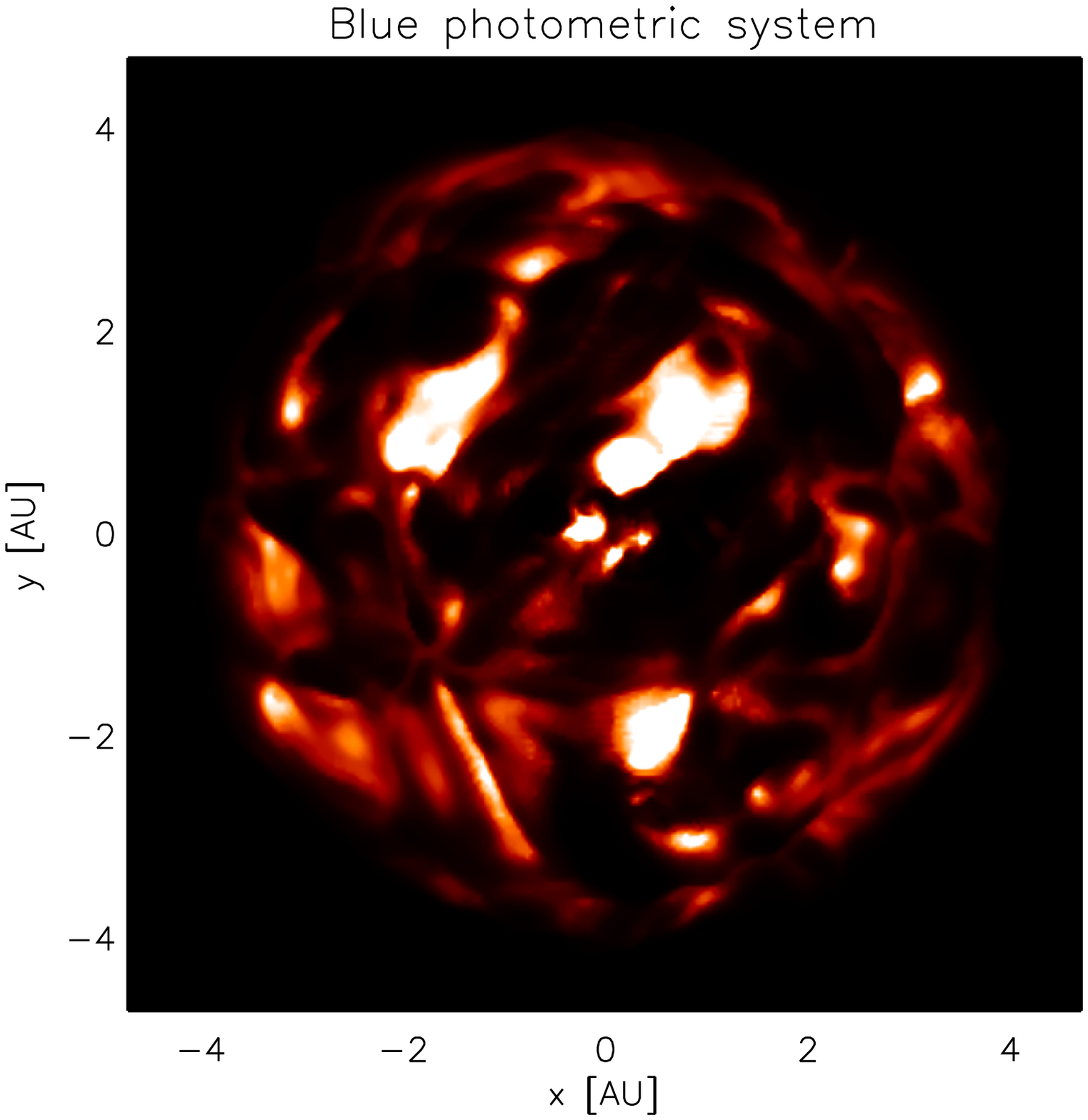}
   \includegraphics[width=0.35\hsize]{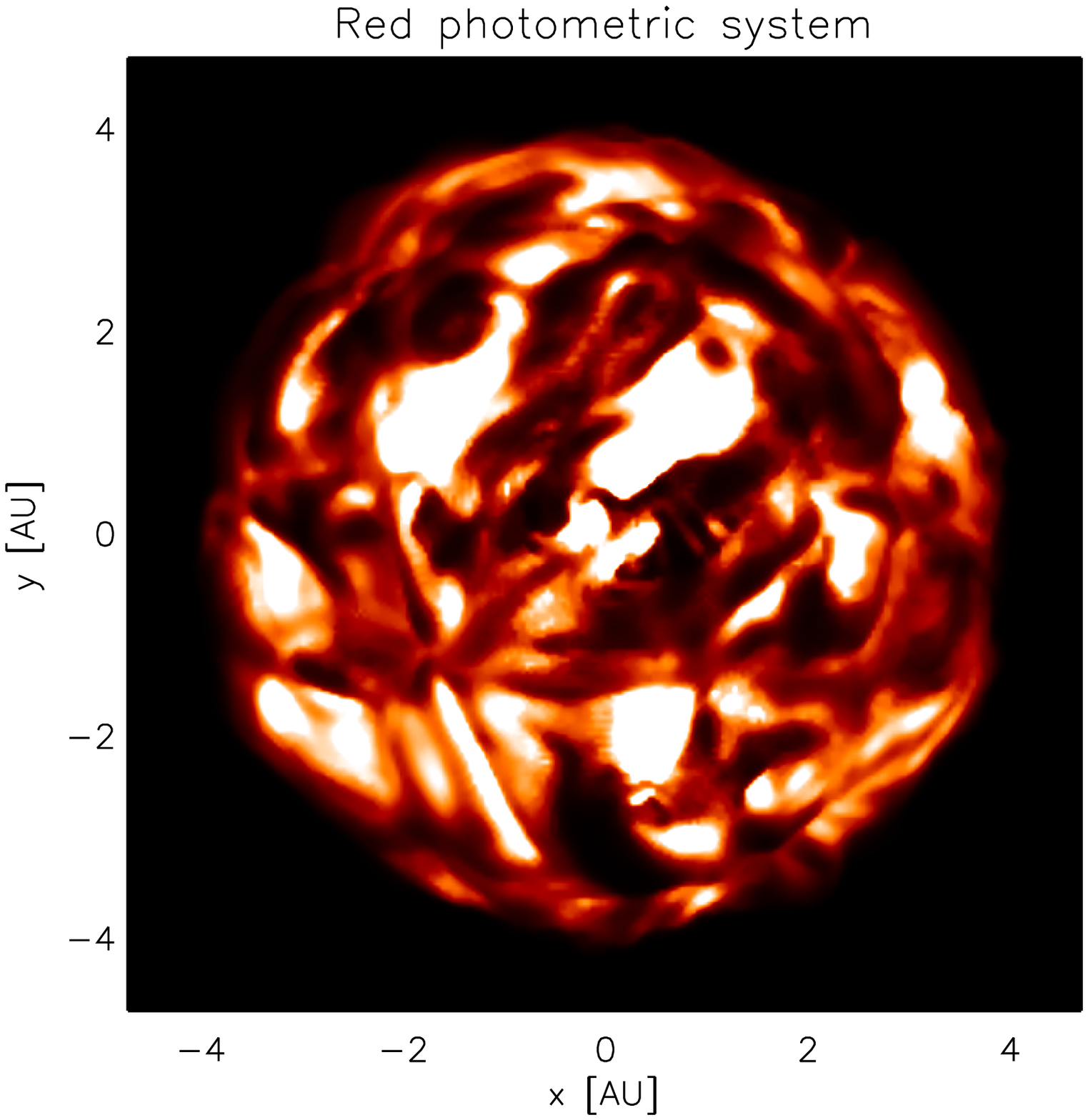}\\
   \includegraphics[width=0.4\hsize]{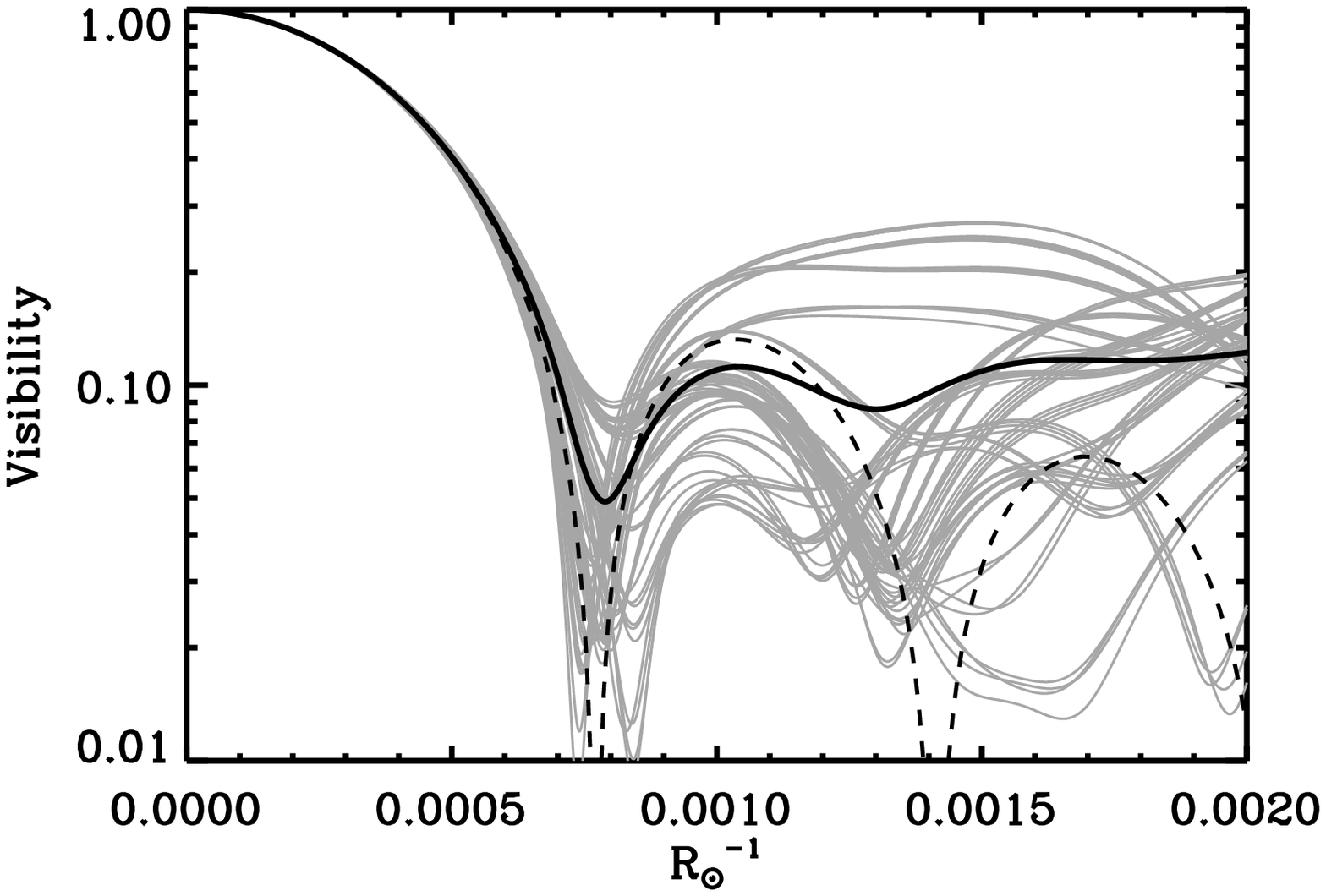}
   \includegraphics[width=0.4\hsize]{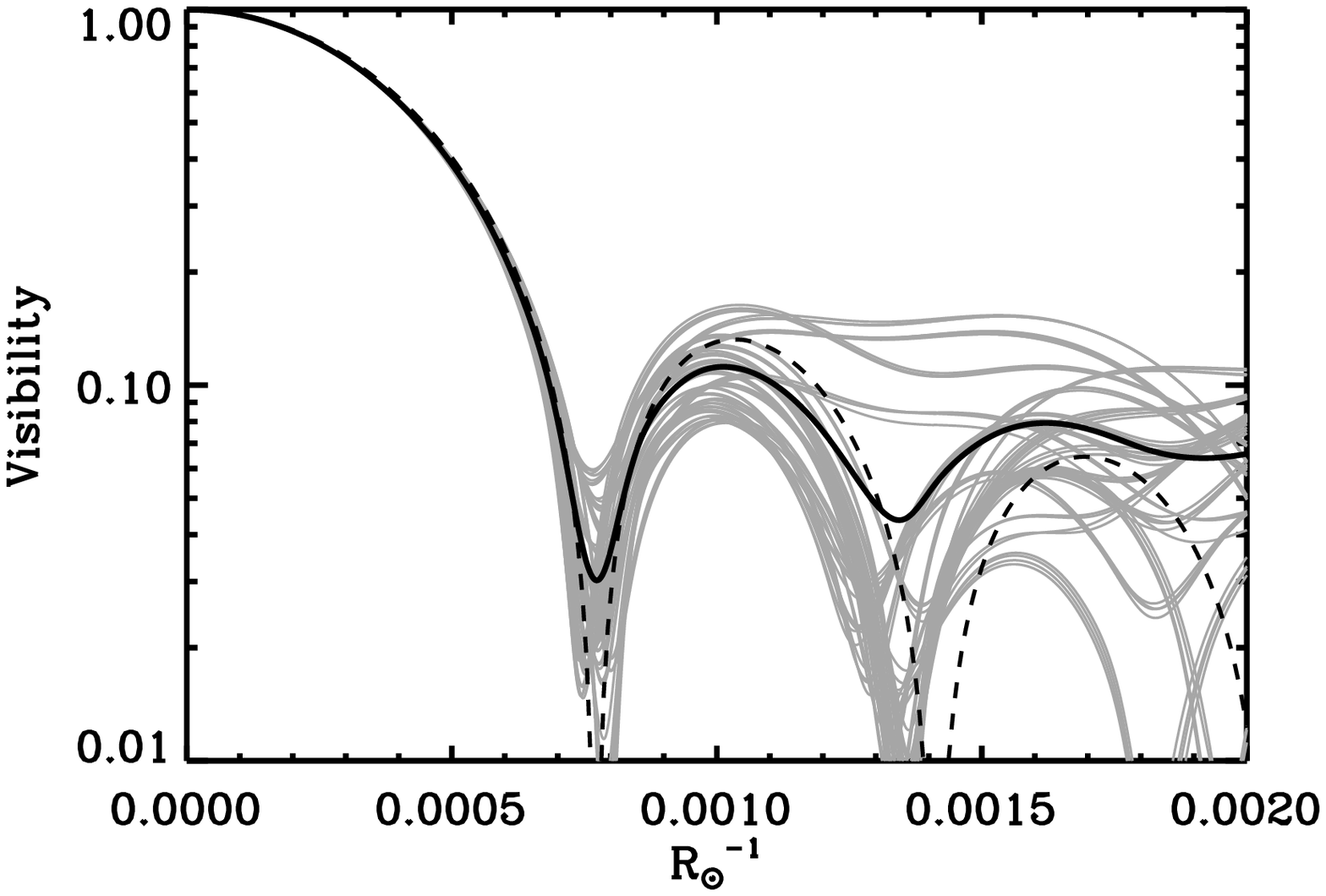}\\
\end{tabular}
      \caption{\emph{Top panels:} maps of the linear intensity (the range is [0 -- 230000]
        erg/s/cm$^2$/\AA ) computed in the blue and red photometric filters of Fig.~\ref{fig1}. \emph{Bottom panels:} visibility curves from the above
maps computed for 36 different position angles 5$^\circ$ apart (grey
lines). The black curve is the average visibility while the dashed
line is a uniform disk of about the same radius as the simulation
snapshot. The conversion
factor to the more customary unit arcsec$^{-1}$ on the abscissa axis is arcsec$^{-1}={\rm R}_\odot^{-1}\cdot d~[{\rm pc}]\cdot214.9$ (see Paper~I).     }
         \label{fig_foto2}
   \end{figure*}

The simulation presented in this work has already been tested against
the observations at different wavelengths including the optical region
(Papers~I and II). However, it is now also possible to compare the
predictions in the Gaia $G$ band to CHARA interferometric observations
obtained with the new instrument VEGA \citep{2009A&A...508.1073M} integrated within the CHARA array at Mount Wilson Observatory. For
this purpose, we computed intensity maps in the blue and red bands of
Fig.~\ref{fig1} and calculated visibility curves for 36 different position angles with a step of
5$^\circ$ following the method explained in
Paper~I. Figure~\ref{fig_foto2} shows the intensity maps together with
the corresponding visibility curves. The angular visibility
fluctuations are larger in the blue band (bottom left panel) because
there is a larger contribution from molecular opacities (mainly TiO) that shade
the continuum brightness of the star (top left panel): therefore the surface brightness contrast is higher. However, in both photometric bands the signal in the second lobe, at higher frequencies, is $\sim\!\!0.2$\,dex larger than the uniform disk (UD) result, which is measurable with CHARA. \\
\indent The approach we suggest to follow in order to check the reliability of the 3D simulation is the following: to search for angular visibility variations, as a function of wavelength, observing with the same telescope configuration covering high spatial frequencies and using the Earth's rotation to study 6-7 different position angles in one night. The error bar should be kept smaller than the predicted fluctuations: $\approx$40$\%$ in the blue band, and $\approx$20$\%$ in the red band at the peak of the second visibility lobe.

\begin{figure}
   \centering
   \includegraphics[width=0.7\hsize]{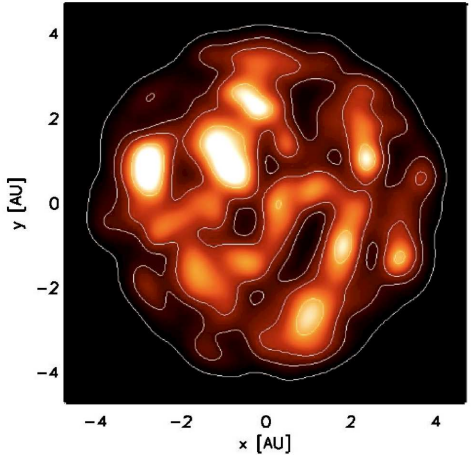}
      \caption{Snapshot of 3D simulation (upper left of Fig.~\ref{fig2}) convolved to the PSF of a 42\,m
        telescope (for a star at a distance of 152.4 pc, see
        solution \#2 in Table~\ref{Tab:parallax}) like the European Extremely Large
        Telescope.
}
         \label{ELT}
   \end{figure}

To investigate the behavior of the local flux fluctuations,
closure phases shall bring invaluable information on the asymmetry of
the source. However, the final consistency check will be an image
reconstruction to compare directly the granulation size and shape, and
the intensity contrast, provided by the planned second generation recombiner of
the VLTI and CHARA optical interferometry arrays. The European
Extremely Large Telescope (E-ELT, planned to be operating in 2018)
with a mirror size 5 times larger than a single VLT Unit Telescope
will be capable of near IR observations of surface details on RSGs
(Fig.~\ref{ELT}).

\section{Impact of photocentric noise on astrometric measurements}\label{Sect:Photocentervariability}

The basic operating mode of astrometric satellites like Hipparcos or Gaia  is to scan the sky and to obtain along-scan\footnote{Across-scan (AC) measurements will be obtained as well by Gaia but will have a lower precision.} positions $\eta_{AL}$, as it was already briefly sketched in Sect.~\ref{sect:photocenter}. The core astrometric data analysis then consists in solving a least-square problem (for the sake of simplicity, we neglect the AC term) \citep{2010IAUS..261..296L}
\begin{equation}
\label{Eq:xi1}
{\mathrm{\large min}\atop\bf{p,a}}\;\; \left(\sum_{i}^{N_{\mathrm{transit}}}\frac{\left[\eta_{i} - \eta(\mathbf{p},\mathbf{a};t_{i})\right]^2}{\sigma^2_{\eta_{i}}}\right)
\end{equation}
for the astrometric parameters $\bf{p}$, and the set of satellite attitude parameters $\bf{a}$, given the $N_{\mathrm{transit}}$ along-scan positions  $\eta_{i}$ at times
 $t_{i}$, the model predictions $\eta(\mathbf{p,a};t_{i})$, and the formal error $\sigma_{\eta_{i}}$ on the along-scan position  $\eta_{i}$ (including centroiding errors and errors due to imperfect calibration or imperfectly known satellite attitude for instance). If $\eta_{i}$ is affected by some supplementary noise coming from the photocentric motion (which is not going to be included in  $\sigma_{\eta_{i}}$), then this photocentric noise of variance $\sigma^2_{P_{\theta}}$ will degrade the goodness-of-fit in a significant manner, provided that  $\sigma_{P_{\theta}} \ga \sigma_{\eta}$. This statement is easily demonstrated from Eq.~(\ref{Eq:xi1}), by writing $\eta_{i} = \tilde{\eta}_{i} + P_{\theta_i}$,
with the first term $\tilde{\eta}_{i}$ representing the astrometric motion, and the second term representing the along-scan photocentric shift:
\begin{equation}
\label{Eq:xi2}
\chi^2 \equiv  \sum_{i=1}^{N_{\mathrm{transit}}} \frac{\left[\tilde{\eta_{i}} + P_{\theta_i} - \eta(\mathbf{p},\mathbf{a};t_{i})\right]^2}{\sigma^2_{\eta_{i}}}
\end{equation}

For the sake of simplicity, we will assume in the following that $\sigma_{\eta}$ is  the same for  all measurements. 
The above equation may be further simplified in the case 
where there  is no correlation between the astrometric  and photocentric shifts, so that the cross-product term  $\sum_{k=1}^{N_{\mathrm{transit}}}  (\tilde{\eta_{k}}  - \eta(\mathbf{p},\mathbf{a};t_{k}))\; P_{\theta_k}$ is null. This absence of correlation only holds if the photocentric shift occurs on time scales 
different from 1 year (no correlation with the parallax), and shorter than a few years (no correlation with the proper motion\footnote{The large subphotospheric convective cells lead to conspicuous spots in the infrared bands, which move on time scales of several years (Paper~I). However, in the optical bands, these large spots are not so clearly visible, since they are swamped in 
smaller-scale photospheric structures.}). Although this assumption of absence of correlation turns out not to be satisfied in real cases (we will return to this issue in the discussion of Fig.~\ref{Fig:parallax}),        it nevertheless offers insights into the situation, and we therefore pursue the analytical developments by writing    
\begin{eqnarray}
\label{Eq:xi3}
\chi^ 2 
 & = & \frac{1}{\sigma^2_{\eta}} \left[\sum_{i=1}^{N_{\mathrm{transit}}}  \left[\tilde{\eta}_{i} - \eta(\mathbf{p},\mathbf{a};t_{i})\right]^2 + \sum_{i=1}^{N_{\mathrm{transit}}}   P_{\theta_i}^2\right] \nonumber\\
& = & \chi^ 2_0 + N_{\mathrm{transit}} \;  \frac{\sigma^2_{P_{\theta}}}{\sigma^2_{\eta}}
\end{eqnarray}
where $\chi^ 2_0$ is the chi-square obtained in the absence of photocentric motion,  
and we have assumed $\sigma^2_{P_{\theta}} = (1/N_{\mathrm{transit}}) \sum_{k=1}^{N_{\mathrm{transit}}}   P_{\theta_k}^ 2$ since 
asymptotically $\langle P_{\theta}\rangle = 0$.  It is important to stress here that it is indeed the standard deviation of the 
photocenter displacement ({\it sampled the same way as the astrometric data have been}) -- rather than its  average value --  which matters.
In the extreme case where there is a constant (non-zero) photocenter shift,  
there will obviously be no impact on the astrometric parameters.

The degradation of the fit due to the presence of the photocentric noise may be 
quantified through the goodness-of-fit parameter  $F2$, defined as 
\begin{equation}\label{Eq:F2}
F2 =  \left(\frac{9\nu}{2}\right)^{1/2} \left[\left(\frac{\chi^2}{\nu}\right)^{1/3}+\frac{2}{9\nu}-1\right],
\end{equation}
where $\nu$ is the number of degrees of freedom 
of the $\chi^2$ variable. The above definition corresponds to  the 'cube-root transformation' of the $\chi^2$ variable \citep{Kendall-1977}. The transformation of ($\chi^2,\nu$) to $F2$ eliminates the inconvenience of having the distribution depending on the additional variable $\nu$, which is not the same for the different stars. $F2$ follows a normal distribution with zero mean and unit
standard deviation. The goodness-of-fit $F2$ thus appears to be an efficient way to detect the presence of any photocentric noise. It may be compared to its value $F2_0$ in the absence of photocentric noise by  assuming $\chi^2_0/\nu = 1$ and $N_{\mathrm{transit}}/\nu \approx 1$; then Eq.~(\ref{Eq:xi3}) writes
\begin{equation}
\chi^ 2 \approx \chi^ 2_0 \left( 1 + \frac{N_{\mathrm{transit}}}{\nu} \;  \frac{\sigma^2_{P_{\theta}}}{\sigma^2_{\eta}} \right),
\end{equation}
thus leading to
\begin{equation}
\label{Eq:DeltaF2}
F2 =  F2_0 + \left(\frac{9\nu}{2}\right)^{1/2}\; \left[ \left(1+ \frac{\sigma^2_{P_{\theta}}}{\sigma^2_{\eta}}\right)^{1/3}  -1 \right].
\end{equation}

\begin{figure}
   \centering
   \includegraphics[width=1.1\hsize]{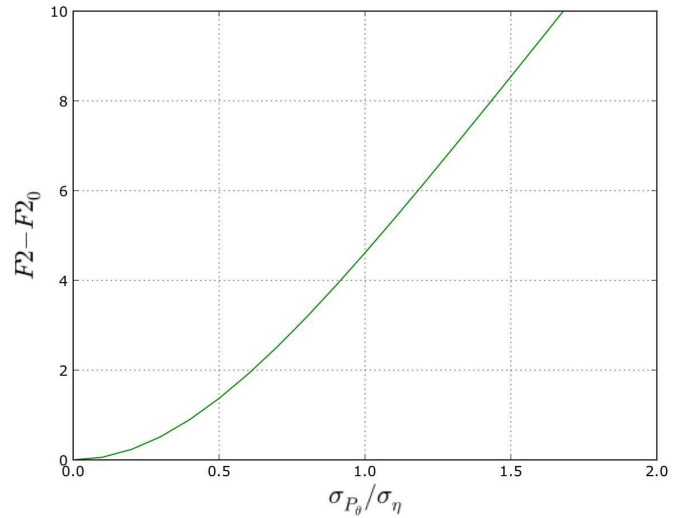}
      \caption{The degradation of the goodness-of-fit $F2$ in the presence of a photocentric motion described in terms of the ratio $\sigma_{P_{\theta}} / \sigma_{\eta}$, where $\sigma_{\eta}$ is the instrumental error. Remember that $F2$ follows a normal distribution with zero mean and unit
standard deviation. 
}
         \label{Fig:sigmaPsigeta}
   \end{figure}

In the case of Gaia, the second term of the above equation may be evaluated as a function of $\sigma_{P_{\theta}}/\sigma_{\eta}$ by adopting $\nu = 70$, as represented on Fig.~\ref{Fig:sigmaPsigeta}. Since $F2$ follows a normal distribution with zero mean and unit
standard deviation, the fit degradation will become noticeable if $F2$ increases by 2 or so, implying $\sigma_{P_{\theta}}/\sigma_{\eta} \ga 0.6$.
This translates into a condition on the distance:
\begin{equation}
\label{Eq:d}
d\; [\mbox{kpc}] \la \frac{\sigma_{P_{\theta}} \; [\mbox{AU}]} {0.6\; \sigma_{\eta}\; [\mbox{mas}]}.
\label{dPw}
\end{equation}
The error on the along-scan position $\eta$ should not be confused 
with the end-of-mission error on the parallax ($\sigma_\varpi$), which ultimately results from the combination of 
$N_{\mathrm{transit}}$ transits, with  $N_{\mathrm{transit}}$ ranging from 59 to 120 for Gaia, with an average of $\bar{N}_{\mathrm{transit}} = 78$  \citep{2010IAUS..261..296L}, and from 10 to 75 for Hipparcos (Fig.~3.2.4 of Vol. 1 of the Hipparcos and Tycho Catalogues). The number of transits depends (mostly) on the ecliptic latitude. 

For Hipparcos, the individual $\sigma_\eta$ values for each transit may be found in the Astrometric Data files \citep{vanLeeuwen-1998:a,vanLeeuwen-2007:a}, and are of the
order of 1.7~mas for the brightest stars (see Sect.~\ref{Sect:Hipparcos} and Fig.~\ref{fig:Phase}). 
For Gaia, the quantity $\sigma_{\eta}$ may be obtained from the relation
\begin{equation}
\sigma_{\eta} = \frac{ N^{1/2}_{\mathrm{transit}}}{ m \ts g_{\mathrm par}} \; \sigma_\varpi,
\end{equation}
where $m = 1.2$ denotes an overall end-of-mission contingency margin, and $g_{\mathrm{par}} = 1.91$ is a dimensionless geometrical factor depending on the scanning law, and accounting for the variation of $N_{\mathrm{transit}}$ across the sky, since $\sigma_{\varpi}$ is an effective sky-average value \citep[see ][]{2005ESASP.576...35D}.
A current estimate of  $\sigma_{\varpi}$ is 7.8~$\mu$as for the brightest stars \citep{2010IAUS..261..296L}, yielding $\sigma_{\eta}$ of the order of 30~$\mu$as.  
To avoid saturation on objects brighter than $G = 12.6$, a special CCD gating strategy will be implemented so that the error budget may be assumed to be a constant for $G \le 12.6$
\citep{2005ESASP.576...35D,2010IAUS..261..296L}.
As we show in Sect.~\ref{Sect:Gaia}, only the bright-star regime matters for our purpose. Inserting these values in Eq.~(\ref{Eq:d}), we thus find
\begin{equation}
\label{Eq:M-d2-Hip}
d\; [\mbox{kpc}]  \le 0.98 \; \sigma_{P_\theta} \; [\mbox{AU}] \hfill\mathrm{for\; Hipparcos,}\;\;\; 
\end{equation}
and
\begin{equation}
\label{Eq:M-d2-Gaia}
d \; [\mbox{kpc}] \le 55.5 \; \sigma_{P_\theta} \; [\mbox{AU}]\hfill\mathrm {for\; Gaia.}\;\;\; 
\end{equation}
Adopting  $\sigma_{P_\theta} = 0.08$~AU 
for Betelgeuse-like supergiants  (Sect.~\ref{sect:photocenter}) yields 
$d< 0.08$~kpc for Hipparcos and $d< 4.4$~kpc for Gaia.
This limit has to be interpreted as marking the maximum distance up to which a photocentric motion with 
$\sigma_{P_\theta} = 0.08$~AU will  increase the astrometry goodness-of-fit by 2. 
The validity of these conditions will be further evaluated  in Sects.~\ref{Sect:Hipparcos} and \ref{Sect:Gaia}.

In the presence of such photocentric noise, the astrometric data reduction process may adopt one of the following three approaches:
\begin{itemize}
\item[(i)] Neither the model definition, nor the measurement-error definition are modified (meaning that the quantities $\sigma_i$ entering Eq.~(\ref{Eq:xi3}) are the same as before, and that no attempt is made whatsoever to model the granulation). With respect to a star with similar properties (same apparent magnitude and location on the sky), a star with global-scale convection cells will then be recognized by a goodness-of-fit $F2$ value larger than expected depending upon the ratio  $\sigma_{P_{\theta}} / \sigma_{\eta}$ (see Fig.~\ref{Fig:sigmaPsigeta}). Under those conditions, the resulting {\it formal} uncertainty on the parallax would {\it not} be especially large, though; actually, it would be exactly identical to the parallax uncertainty in the absence of photocentric motion.  This is because the formal errors 
on the parameters $\mathbf{p}$ (among which the parallax) only depend on the measurement errors $\sigma_{\eta}$  (which were kept the same in the presence or absence of photocentric noise), and not on the actual measured values (which will be different in the two situations).  This is demonstrated in Appendix~A.  
But of course, the error on the parallax derived in such a way is underestimated, as it does not include the extra source of noise introduced by the photocentric motion. The next possibility alleviates this difficulty. 
\item[(ii)] An estimate of the  photocentric noise may be quadratically added  to $\sigma_{\eta}^ 2$ appearing in Eq.~(\ref{Eq:xi2}). The error on the parallax will then be correctly estimated (and will be larger than the one applying to similar non-convective stars); the goodness-of-fit will no longer be unusually large. 
This is the method adopted for the so-called 'stochastic solutions' in the Hipparcos reduction, an example of which will be presented in Sect.~\ref{Sect:Hipparcos}. These solutions, called 'DMSA/X', added some extra-noise (in the present case: the photocentric noise) on the measurements to get an acceptable fit.
\item[(iii)]  The model is modified to include the photocentric motion. This would be the best solution in principle, as it would allow to alleviate any possible error on the parallax, as they may occur with the two solutions above. However, the 3D simulations reveal that it is very difficult to model the complex convective features seen in visible photometric bands by a small number of spots with a smooth time behaviour.  This solution has thus not been attempted.
\end{itemize} 

The astrometric parameters themselves may change of course, for either of the above solution, especially when the photocentric motion adds to the parallactic motion a signal having a characteristic time scale close to 1~yr. If on the contrary, the photocentric motion has a characteristic time scale very different from 1~yr, the photocentric motion averages out, and leaves no imprint on the parallax. A similar situation is encountered in the presence of an unrecognized orbital motion on top of the parallactic motion: only if the unaccounted orbital signal has a period close to 1~yr will the parallax be strongly affected \citep[see ][for a discussion of specific cases]{2000A&AS..145..161P}.  As explained in Section~\ref{sect:photocenter}, RSGs large convective cells evolve over time scales of years. In addition, they change slightly their position on the stellar surface within the 5 years of simulated time (Chiavassa Ph.D. thesis\footnote{http://tel.archives-ouvertes.fr/docs/00/29/10/74/PDF/Chiavassa\_ PhD.pdf}), but, unfortunately, it is difficult to measure exactly the granule size (Berger et al. 2010, submitted to A$\&$A) and thus to give a consistent estimation of this displacement.

\section{A look at Betelgeuse's Hipparcos parallax} \label{Sect:Hipparcos}

\begin{figure}
  \centering
    \begin{tabular}{cc}
	\includegraphics[width=0.45\textwidth]{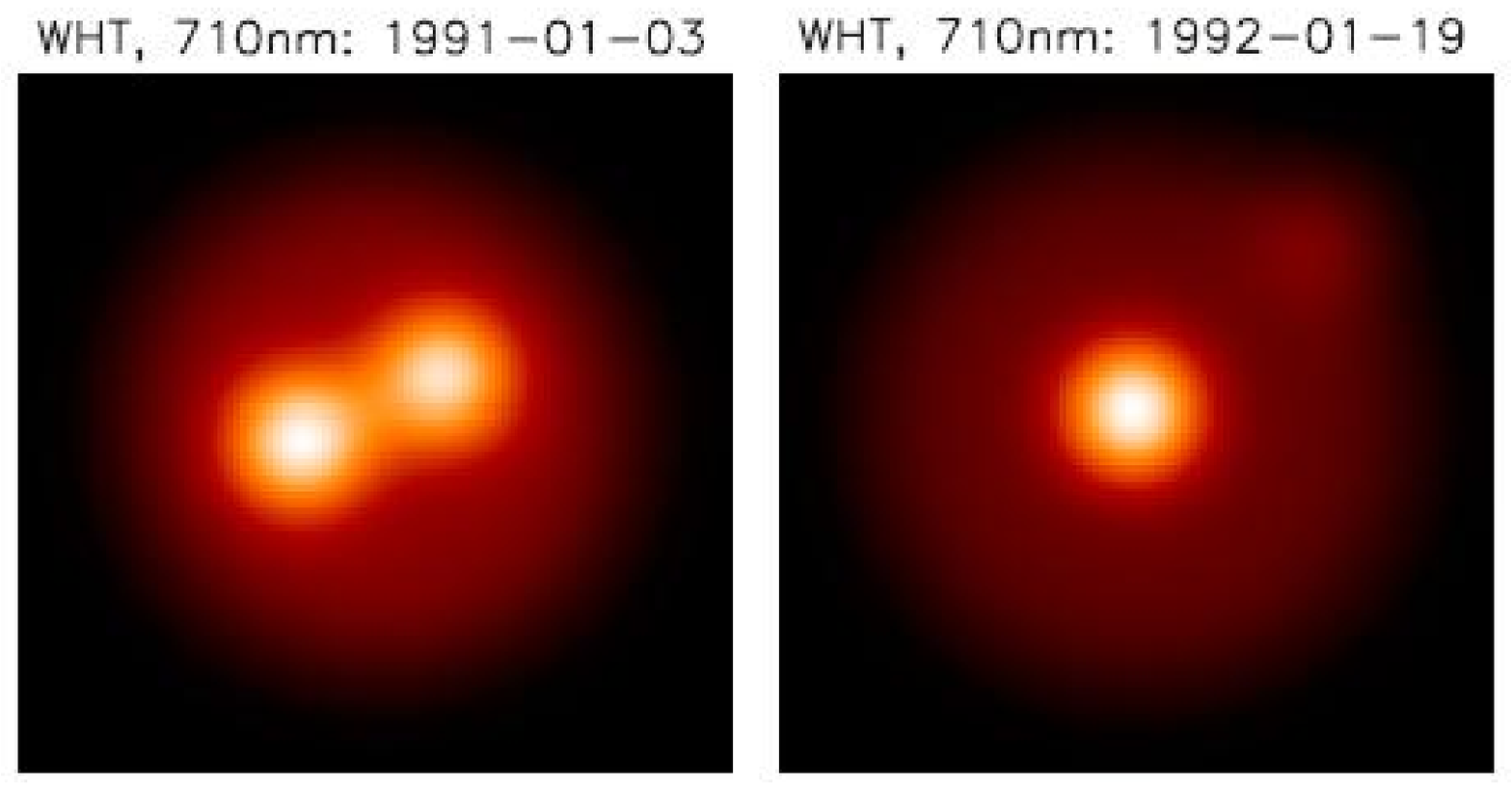}\\
	 \includegraphics[width=0.45\textwidth]{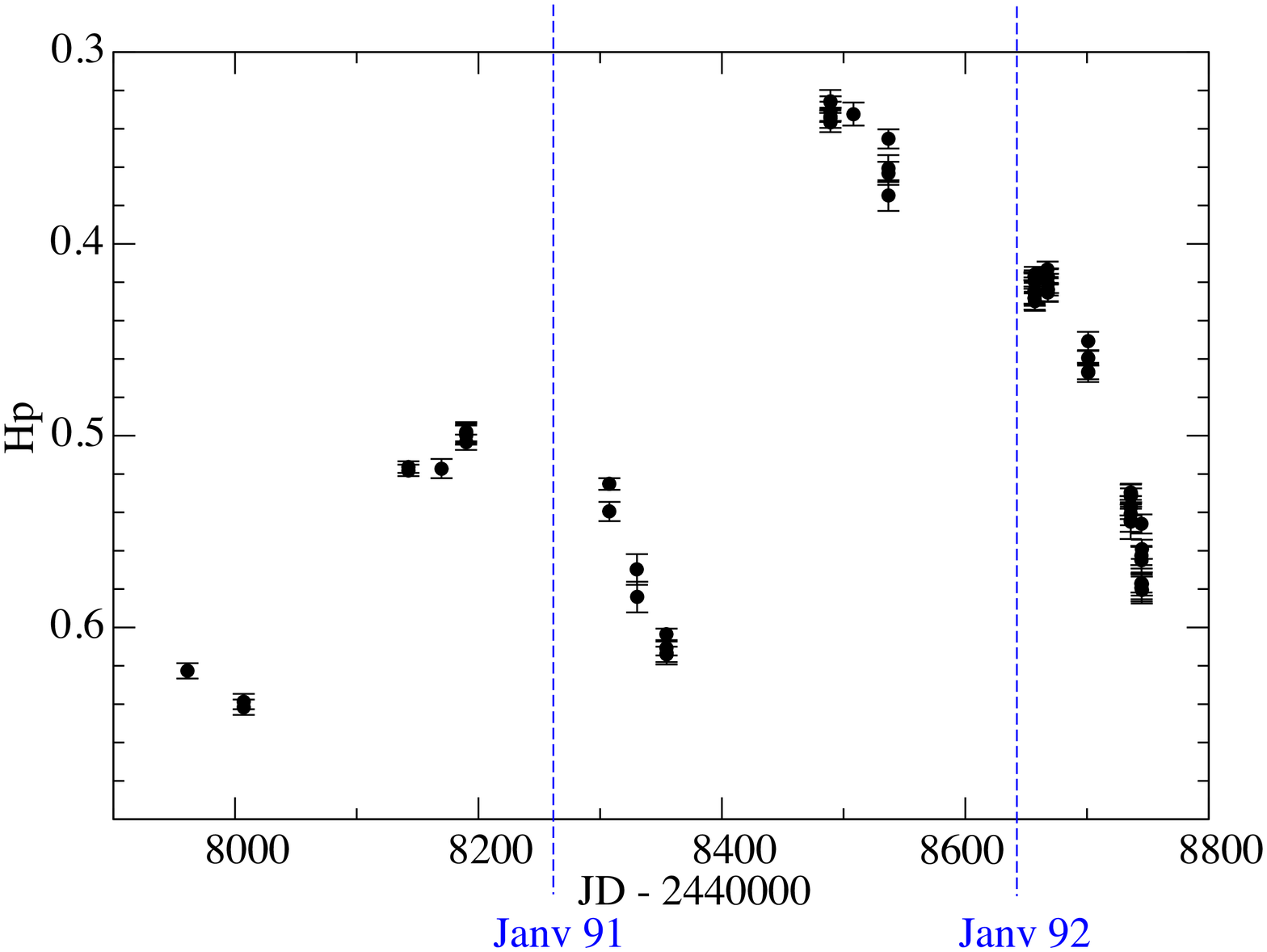}
  \end{tabular}
  \caption{\emph{Top panel:} interferometric observations of the
  surface of Betelgeuse obtained with the William Herschel Telescope at a
  wavelength of 710~nm, in January 1991 \citep{1992MNRAS.257..369W} and
  January 1992 \citep{Tuthill-1997}. North is on top and East is to
  the left. The two images have been taken from
  \cite{2002AN....323..213F}. The spot properties are summarised in
  Table~\protect\ref{Tab:Betelgeuse}. 
  \emph{Bottom panel:} Hipparcos epoch photometry of Betelgeuse with the
  vertical lines indicating the epoch of the two
  interferometric observations.} 
  \label{fig:BetWHT}
\end{figure}

Hipparcos data \citep{ESA-1997} may  hold  signatures of
global-scale granulation in supergiants. The three nearby supergiants
$\alpha$~Sco (Antares; HIP~80763), $\alpha$~Ori (Betelgeuse; HIP~27989) and $\alpha$~Her (Rasalgethi; HIP~84345) are
ideal targets for this purpose, since \citet{Tuthill-1997} indeed found
surface features on all three stars, implying photocentric displacements of
the order of 1~mas (estimated from the product of the fraction of
flux belonging to a bright spot with its radial distance from the
geometric center; see Table~\ref{Tab:Betelgeuse}). By chance, observations of the disc of Betelgeuse at the
time of the Hipparcos mission were done by
\citet{1992MNRAS.257..369W} and \citet{Tuthill-1997} and are shown in
Fig.~\ref{fig:BetWHT}. \citeauthor{Tuthill-1997} reveal that the two bright spots present in
January~1991 turned into one a year later (January~1992), with a much
fainter spot appearing at the edge of the extended disc. Since the simulation used in this work shows excellent fits to the visibility curves, closure phases, and reconstructed images based on WHT data in the same filters as those used in Fig.~\ref{fig:BetWHT}, and the fact that RSGs are slow rotators, it is most likely that the spots in Fig.~\ref{fig:BetWHT} are due to convection. Their
properties have been summarised in Table~\ref{Tab:Betelgeuse}, along with the
corresponding photocentric displacement. The observed photocentric displacements agree with the model predictions, as can be evaluated from 
$\sigma_{P_\theta} \mathrm{[mas]} = \sigma_{P_\theta} \mathrm{[AU]} \times \varpi \mathrm{[mas]} = 0.5$~mas, 
since $\sigma_{P_\theta} = 0.08$~AU for a
supergiant (Sect.~\ref{Sect:Photocentervariability}) and $\varpi =
6.56\pm0.83$~mas for Betelgeuse (see Table~\ref{Tab:parallax}; in the
remainder of this section, all quantities from Hipparcos refers to van
Leeuwen's  new reduction, \citeyear{vanLeeuwen-2007}).

\begin{table*}
\caption{\label{Tab:Betelgeuse}
Properties of the spots observed at the surface of Betelgeuse during
the Hipparcos mission. The position offset refers to the center of the
extended disc, of radius 27~mas (Jan.~1991) and 23~mas (Jan.~1992). 
Data from \citet{1992MNRAS.257..369W,Tuthill-1997}.
}
  \centering
  \begin{tabular}{l|c|c|c|c|c|c|cc}
   & \multicolumn{3}{c|}{Spot 1} & \multicolumn{3}{|c}{Spot 2} &  \multicolumn{2}{|c}{\bf Photocenter}\\
   & pos. offset & pos. angle & flux fraction & pos. offset & pos. angle & flux fraction & pos. offset & pos. angle\\
   & [mas] & [$^\circ$] & \% & [mas] & [$^\circ$] & \% & [mas] & [$^\circ$]\\
  \hline
  January 1991 & 9 $\pm$ 2& 105$\pm$3 & 12 $\pm$ 2 &  9 $\pm$ 2 &  305$\pm$4 & 11 $\pm$ 2 &{\bf 0.4}
  & {\bf 39}\\
  January 1992 & 2 $\pm$ 1& 40$\pm$10 & 17 $\pm$ 2 & 29 $\pm$ 3 &
   -45$\pm$5 & 4 $\pm$ 1 & {\bf 1.2} & {\bf -29}\\
\hline
  \end{tabular}
\end{table*}

Considering the fact that the instrumental uncertainty on an
individual measurement is 1.9~mas for a star like Vega  ($V = 0.03$) which is as bright as Betelgeuse ($V = 0.42$)
\citep[according to Vega's Intermediate Astrometric
Data file in ][]{vanLeeuwen-2007:a}, Betelgeuse's convective noise with $\sigma_{P_\theta} = 0.5$~mas should be just noticeable on top of the instrumental noise, and possibly have some detectable impact  on the
astrometric data of Betelgeuse. For Vega, \citet{vanLeeuwen-2007:a} found a very good astrometric solution whose residuals $\Delta\eta$ have a standard deviation $\sigma_{\Delta\eta}$ of 1.78~mas, fully consistent with the formal errors on $\eta$ (top panel of Fig.~\ref{fig:Phase}). The extreme brightness of Vega thus did not prevent from finding a good astrometric solution. 
On the other hand, neither the original Hipparcos processing nor  van Leeuwen's revised processing could find an acceptable fit to Betelgeuse and Antares astrometric data, and a so-called 
'stochastic solution' (DMSA/X) had to be adopted (the kind of solution labelled (ii) in the discussion of Sect.~\ref{Sect:Photocentervariability}), meaning that some supplementary noise  (called 'cosmic noise') had to be added  to yield acceptable goodness-of-fit values $F2$.  

\begin{figure}
\centering
\includegraphics[width=0.45\textwidth]{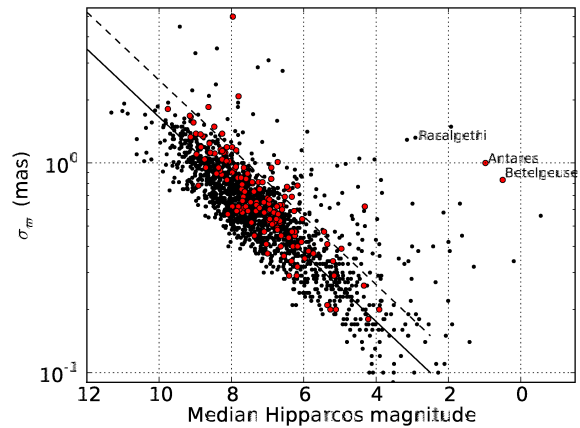}
\caption{Parallax standard errors for supergiants in the
  Hipparcos Catalogue, after van Leeuwen's reprocessing. 
The dashed line represents 1.5 times the standard parallax error 
for  stars with precise parallaxes ($<10$\%; solid line) in van Leeuwen's reprocessing  \citep[see Fig. 2.19 of][]{vanLeeuwen-2007:a}, whereas the solid line represents the fiducial relation between the Hipparcos magnitude and the standard parallax error.
Large red circles correspond to stars redder than $V-I=2$.} 
\label{fig:HpsigpiSup}
\end{figure}

\begin{figure}
\centering
\includegraphics[width=1.05\hsize]{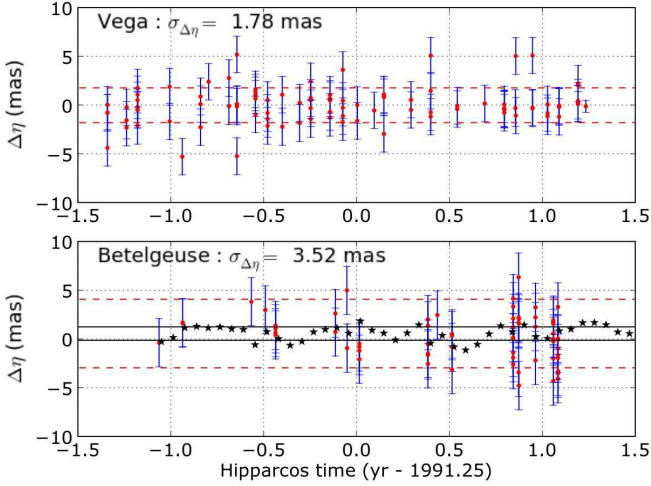}
\caption{Abscissa residuals $\Delta \eta$ (in mas on the sky; red dots) along with the corresponding error bar
from \citet{vanLeeuwen-2007:a}
 for Betelgeuse and Vega, as a function of time,
expressed in years from 1991.25.  
Filled  star symbols correspond to the along-scan projections $P_\theta$ of the synthetic photocenter displacements  of Fig.~\ref{fig3}.  The red dashed lines and black solid lines depict the $\pm 1 \sigma$ interval around the mean for the Hipparcos data points and model predictions, respectively. 
Note that these displacements were computed in the Gaia $G$ filter instead of the Hipparcos $Hp$ filter. A test on a given snapshot 
has shown that the difference is negligible: $P_x = 0.11$~AU with the $Hp$ filter, as compared to 0.13~AU with the $G$ filter.
}
\label{fig:Phase}
\end{figure}

\begin{figure}
\centering
\includegraphics[width=1.05\hsize]{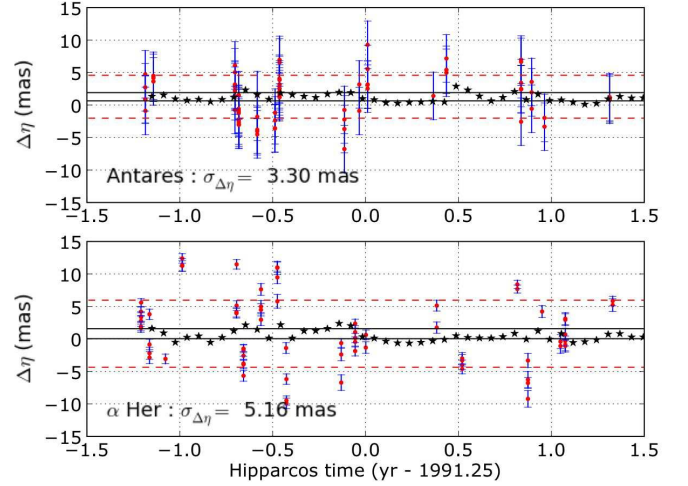}
\caption{Same as Fig.~\ref{fig:Phase} for Antares and $\alpha$ Her.
}
\label{fig:Antares}
\end{figure}

The cosmic noise amounts to 2.4 and 3.6~mas for Betelgeuse
and Antares, respectively, in van Leeuwen's  reprocessing. These values correspond to the size of the error bars displayed on Figs.~\ref{fig:Phase} and \ref{fig:Antares}. Rasalgethi was not flagged as DMSA/X, but rather as DMSA/C
(indicating the presence of a close companion), but its
large goodness-of-fit value $F2 = 46.63$ is indicative as well of increased noise. 
Consequently,  all three supergiants  have a  parallax standard 
error larger than expected\footnote{This larger parallax standard error does not contradict Appendix A stating that, in the presence of a photocentric 
noise, the standard error on the parallax should stay the same. This is because this parallax standard error is obtained 
in the framework of a DMSA/X ("stochastic") solution, where the measurement errors have been artificially 
increased by a "cosmic noise" to get an acceptable goodness-of-fit value. Hence, the "design matrix" defined
in Appendix A, and directly related to the variance-covariance matrix of the astrometric parameters, 
{\it has} been changed to produce the stochastic solution, thus resulting in a larger parallax error. This corresponds to a solution of kind (ii) in Sect.~\ref{Sect:Photocentervariability}.} given its Hipparcos magnitude $Hp$, as revealed by 
Fig.~\ref{fig:HpsigpiSup} which displays $\sigma_\varpi$ against the median magnitude for all 
supergiants (luminosity classes~I and II, of all spectral types) in the Hipparcos catalogue. The chromaticity correction has been a serious concern 
for the reduction of the Hipparcos data of very red stars \citep[see][for a detailed discussion of this problem]{2003A&A...397..997P}, and one may wonder whether the increased noise of the three supergiants under consideration could perhaps be related to this effect. 
Since the very red supergiants (with $V-I > 2.0$) show no appreciable offset  from the rest of the sample in Fig.~\ref{fig:HpsigpiSup} (at least for the brightest supergiants, down to $Hp = 8$), this possibility may be discarded, and the discrepant  behavior of Rasalgethi, Antares and Betelgeuse in Fig.~\ref{fig:HpsigpiSup} seems instead related to their large apparent brightness, due to their proximity to the Sun.

Could  the poor accuracy of Betelgeuse's parallax and its cosmic noise be related
to its surface features, as already suggested in general terms by
\citet{Barthes-Luri-1999,Gray-2000,2003A&A...397..997P,2005ESASP.560..979S,2005ESASP.576..215B,2006A&A...445..661L,Eriksson-2007}.


The bottom panel of Fig.~\ref{fig:Phase} shows the along-scan residuals $\Delta
\eta$ for Betelgeuse against time (and Fig.~\ref{fig:Antares} does the same for Antares and Rasalgethi),
compared with the photocenter displacements $P_x$ and $P_y$ determined from the 3D simulation of Sect.~\ref{sect:photocenter}. 
  
From this comparison, we conclude that the photocentric noise, as predicted by the 3D simulations, does account for a substantial part of the 'cosmic noise', but not for all of it. 
A possibility to reconcile predictions and observations could come from an increase of Betelgeuse's parallax (because the observed photocentric motion would then be larger for the $\sigma_{P_\theta}$ value fixed by the models), but this suggestion is not borne out by the recent  attempt to improve upon Betelgeuse's
parallax in the recent literature \citep{2008AJ....135.1430H} (solution \#2 in Table~\ref{Tab:parallax}), by combining the Hipparcos astrometric data with VLA positions, as this new value is  {\it smaller} than both the original Hipparcos and van Leeuwen's values. The remaining possibility is that the 3D model discussed in Sect.~\ref{sect:photocenter}
underestimates the photocentric motion. In fact, Paper~II showed that the RHD simulation fails to reproduce the TiO molecular band strengths in the optical region (see spectrum in Fig.~\ref{fig1}). This is due to the fact that the RHD simulations are constrained by execution time and
    therefore use a grey approximation for the radiative transfer. This is well
    justified in the stellar interior, but is a crude approximation in the optically thin
layers. As a consequence, the thermal gradient is too shallow and weakens
the contrast between strong and weak lines
\citep{2006sf2a.conf..455C}. The resulting intensity maps look sharper than observations (see Paper~II) and thus also the photocenter displacement should be affected. As described in Paper~II, a new generation of non-grey opacities (five wavelength bins employed to describe the wavelength
dependence of radiation fields) simulation is under development. This will change the
mean temperature structure and the temperature fluctuations, especially in the outer layers where TiO absorption occurs.



\begin{table*}
\centering
\caption[]{\label{Tab:parallax}
Parallaxes for Betelgeuse computed from various data sets. 
}
\begin{tabular}{lllrlrllll}
\hline\\
\# & Data & $N$ & $\Delta t$ & $\varpi$ & Distance\\
    &       &     &   [yr]     & [mas] & [pc] \\
\hline\\
1 & Hipparcos (FAST+NDAC reduction consortia) & 38 & 2.1 & $7.63\pm1.64$ & $131.1^{+35.8}_{-23.2}$\\
  \medskip\\
2 & \cite{vanLeeuwen-2007}    & 38 & 2.1 & $6.56\pm0.83$ & $152.4^{+17.1}_{-22.0}$\\
3 & \cite{2008AJ....135.1430H}  & 51 & 22.4& $5.07\pm1.10$ & $197.2^{+35.1}_{-54.6}$ \\
\hline\\
\end{tabular}
\end{table*}





\section{Application to Gaia}\label{Sect:Gaia}

\subsection{Number of supergiant stars with detectable photocentric motion}\label{Sect:Gaia-frequency}

In this section, we will use Eq.~(\ref{Eq:M-d2-Gaia}) to estimate  the number of supergiants which will have a poor goodness-of-fit as a consequence of their 
photocentric motion. This equation requires knowledge of $\sigma_{P_\theta}$, which will be kept as a free parameter in this section. In Sect.~\ref{sect:photocenter},
$\sigma_{P_\theta} = 0.08\,$AU was considered as typical for Betelgeuse-like supergiants, but Sect.~\ref{Sect:Hipparcos} has provided hints that 3D models with grey opacities could somewhat underestimate this quantity. Moreover, according to \citet{2001ASPC..223..785F} and \citet{2006A&A...445..661L}, $\sigma_{P_\theta}$ is expected to vary with the star's atmospheric pressure scale height, which in turn depends upon the star's absolute magnitude $M_G$.    
To explore the parameter space, we thus need to know how $\sigma_{P_\theta}$ varies with $M_G$. This is especially important since on
top of the condition in Eq.~(\ref{Eq:M-d2-Gaia}) relating $d$ to $\sigma_{P_\theta} (M_G)$, there is another constraint coming from the requirement not to saturate the CCD, namely the Gaia magnitude $G$ should be fainter than 5.6. All these constraints may be conveniently encapsulated in boundaries in the $d - M_G$ plane, as displayed in Fig.~\ref{fig_gaia1}.

But first, we have to clarify the relation between  $\sigma_{P_\theta}$ and  $M_G$ which appears to be a critical ingredient in the present discussion. 
Unfortunately, 3D hydrodynamical models in the literature are scarce.
Their main properties are collected in Table~\ref{tab3d}. These
simulations are of two kinds: (i) \emph{box-in-a-star} models cover
only a small section of the surface layers of the deep convection
zone, and the numerical box includes some fixed number of convective cells, large enough to
    not constrain the cells by the horizontal (cyclic) boundaries;  (ii) \emph{star-in-a-box}
models, like the one described in this paper (Sect.~2), cover the
whole convective envelope of the star and have been used to model RSG
and AGB stars so far \citep[see][for an AGB model]{Freytag2008A&A...483..571F}, whereas the former simulations cover a large
number of stellar parameters from white dwarf to red giant stars. The
transition where the box-in-a-star models become inadequate occurs around
$\log g \approx 1$, when the influence of sphericity becomes important; the star-in-a-box global models are then needed, but those are highly computer-time demanding and difficult to run so there are only very few models available so far. 

\cite{2006A&A...445..661L} found that there is a tight correlation between the amplitude of the photocentric motion and the size of the granular cells. This size is related to the pressure scale height at optical-depth unity \citep{2001ASPC..223..785F}. The pressure scale height is defined as
\begin{eqnarray}
\mathcal{H}_{\mathrm{p}}= \frac{k_B T_{\mathrm{eff}}}{m
  g},
\end{eqnarray}
where $g$ is the surface gravity, $k_B$ is the Boltzmann constant and $m$ is the mean molecular
mass ($m=1.31\times m_{\rm{H}} = 1.31 \times 1.67 \times10^{-24}$~grams, for
temperatures lower than 10\ts000~K). In the above expression,  $\mathcal{H}_{\mathrm{p}}$  has the dimension of length. 
But in the remainder of this paper, we adopt instead the simplified definition:
\begin{equation}
\label{eqHp}\label{Eq:HP}
H_{\rm{p}}= \frac{T_{\rm eff}}{
  g}.
\end{equation}

\begin{figure}
\centering
  \begin{tabular}{cc}
   \includegraphics[width=0.65\hsize,angle=-90]{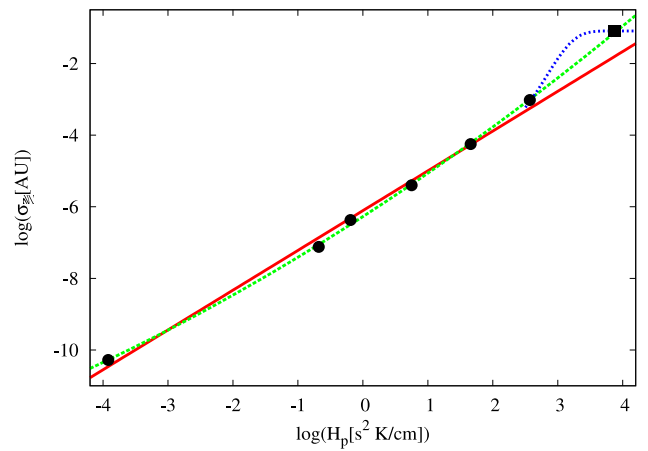}
\end{tabular}
      \caption{Fits to the standard deviation $\sigma_{P_\theta}$ of the photocentric motion predicted from
        3D simulations, as listed in Table~\ref{tab3d}, against the pressure scale height $H_{\rm{p}}$. The equation of the (red) solid line is
        $\log \sigma_{P_\theta}=-6.110+  1.110\; \log(H_{\rm{p}})$ 
with
        $\chi^2=0.17$: the fit considers only the box-in-a-star models (filled circles) of \cite{2005ESASP.560..979S}.
        The (green) dashed line (with equation 
$\log \sigma_{P_\theta} = -6.275+1.174 \log H_{\rm{p}} + 0.039 (\log H_{\rm{p}})^2$)
 is a fit to {\it all} the models of Table~\ref{tab3d} (i.e., box-in-a-star models and star-in-a-box, the latter being represented by a filled square). The (blue) dotted line is
        an arbitrary exponential law that connects the last two points
       with the following equation 
$\log \sigma_{P_\theta} = -1.09 - 3.434 \exp(-0.00149 H_{\rm{p}})$.
}
         \label{fig_gaia2}
   \end{figure}

\begin{figure}
\centering
  \begin{tabular}{cc}
   \includegraphics[width=0.65\hsize,angle=-90]{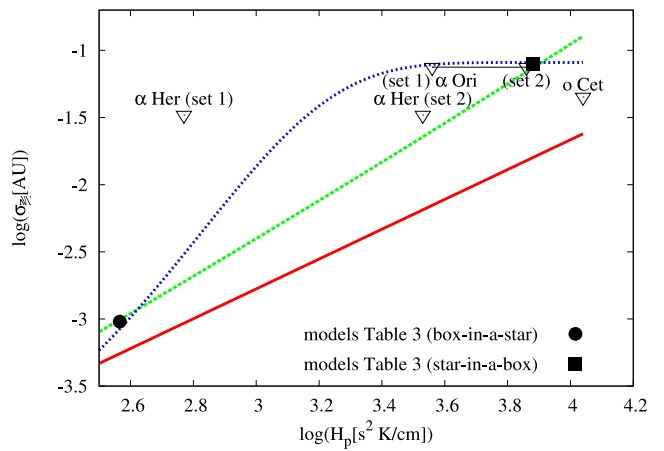}
\end{tabular}
      \caption{Photocenter motions determined from interferometric
        observations for some evolved stars (see Table~\ref{Tab:observP})
        overplotted on the different fits of the standard deviation
        $\sigma_{P_\theta}$ of the photocentric motion as a function of the pressure scale height $H_{\rm{p}}$. The large open inverted triangles 
correspond to the standard deviations of the photocentre deviations for a given observed star. 
Star-in-a-box and box-in-a-star models correspond respectively to the large filled 
square and circle.
      }
   \label{fig_gaia2bis}
   \end{figure}

\begin{figure*}
\centering
  \begin{tabular}{cc}
   \includegraphics[width=0.6\hsize,angle=-90]{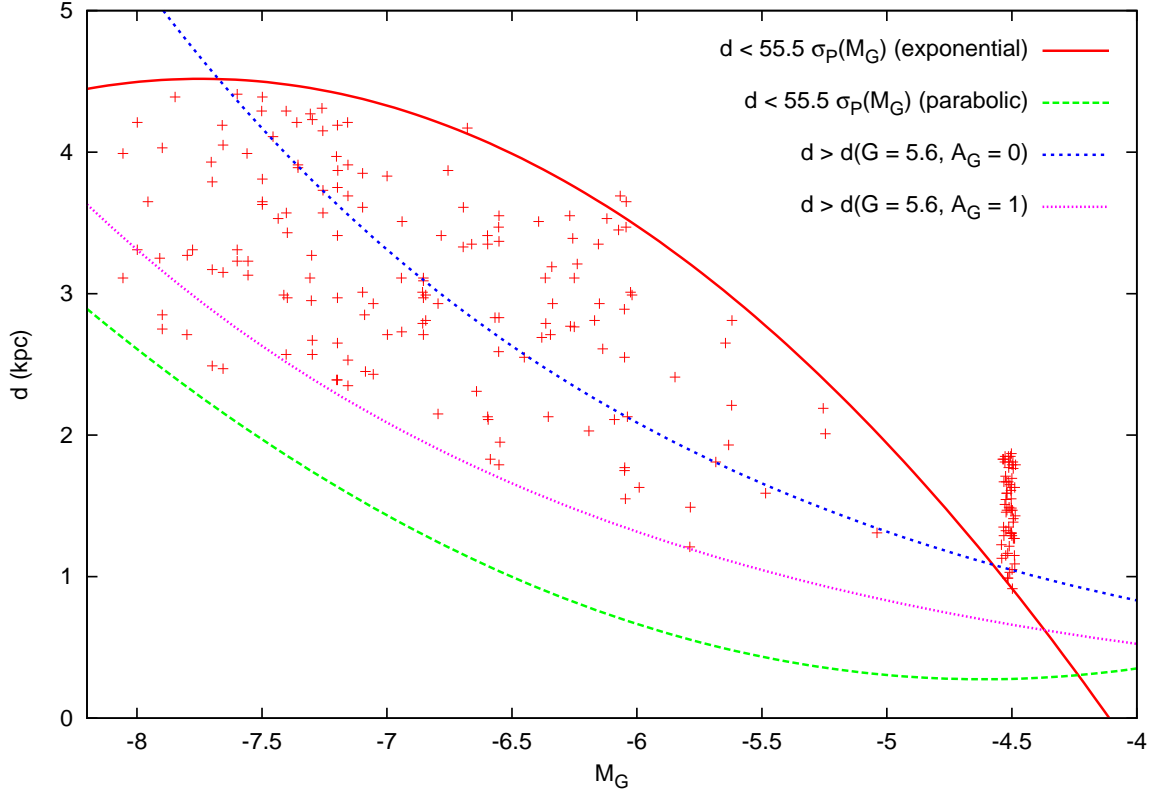}
\end{tabular}
      \caption{
The location in the ($M_G, d$) plane for supergiants with a photocenter noise (of standard deviation $\sigma_{P_\theta}$) 
significantly altering the goodness-of-fit of the astrometric solution (see text), for the case of an exponential link between box-in-a-star and star-in-a-box models (see Fig.~\ref{fig_gaia2bis} and text).  
}
         \label{fig_gaia1}
   \end{figure*}

The law relating the standard deviation of the photocenter displacement to $H_{\rm{p}}$ may be inferred from Fig.~\ref{fig_gaia2} which displays the values from 
Table~\ref{tab3d}. The transition from the most evolved box-in-a-star model (with $\log H_{\rm{p}} \approx  2.57$) to our star-in-a-box model  
($\log H_{\rm{p}} \approx  3.85$) is still unexplored; consequently, there is
no guarantee that the trend obtained at $\log H_{\rm{p}} <  2.57$ may be
extrapolated to larger $H_{\rm{p}}$ values. Different trends are therefore
considered in Fig.~\ref{fig_gaia2} with a zoom in
Fig.~\ref{fig_gaia2bis}. The linear fit of $\log \sigma_{P_\theta}$ as a function of  $\log H_{\rm{p}}$
considers only the box-in-a-star models of \cite{2005ESASP.560..979S}; the parabolic function is the best fit
to all the models (including the star-in-a-box supergiant model). However, there is strong evidence in the simulations that the convective pattern changes strongly from the giant (big black circle symbol in Fig.~\ref{fig_gaia2bis}) to the RSG simulations (big black squared symbol). The convective related surface structures grow enormously in the RSGs and together with the low effective temperature (i.e., the molecular absorption, strongly related to the temperature inhomogenities, is more important) increase the displacement of the photocenter position (i.e., $\sigma_{P_\theta}$ is larger). Thus, the parabolic fit, which considers all the simulations' configurations together, is not a completely correct approach because of the physical changes reported above. Since the transition region between the box-in-a-star (giant stars) and star-in-a-box models (RSG stars) is still unexplored, we consider an extreme transition by  adopting an arbitrary exponential law to relate the last two model simulation points (i.e., the transition region between the box-in-a-star and star-in-a-box models). Paper~I pointed out that the reasons for the peculiar convective pattern in RSGs could be: (i) in RSGs, most of the downdrafts will not grow fast
enough to reach any significant depth before they are swept into
the existing deep and strong downdrafts enhancing the strength
of neighboring downdrafts; (ii) radiative effects and smoothing
of small fluctuations could matter; (iii) sphericity effects and/or numerical resolution (or lack of it). 

To see which among these three possible trends has to be preferred, we have made a compilation of photocentric displacements $P$ from interferometric observations of various supergiants available in the literature (see Fig.~\ref{fig_gaia2bis}). 
Supergiants and Miras have been observed several times in the last
decade with interferometers, often revealing the presence of
surface brightness asymmetries. In several  cases ($\alpha$~Ori, $\alpha$~Her, and  $o$~Cet; see Table~\ref{Tab:observP} for
the data  list and references; more stars will be presented in Sacuto et al., in preparation), 
the observations could be represented by parametric models consisting of a uniform disk
plus one (or more) bright or dark spots.
Using the parameters of the spots fitting the interferometric data, we
computed the positions of the photocenter  for all
observations of a given star and from there the standard deviation of these photocentric positions, which was then plotted against $H_{\rm{p}}$  in Fig.~\ref{fig_gaia2bis}.  
These observational data suggest that the
exponential and quadratic fits of the simulation data are to be
preferred over the linear extrapolation of the box-in-a-star values (Fig.~\ref{fig_gaia2bis}). We stress, however, that the surface gravity
for supergiants like $\alpha$~Her and $\alpha$~Ori are quite uncertain (see Table~\ref{Tab:observP}) and also the highly uncertain metallicity
differences might play a role here.

\begin{figure}
\centering
  \begin{tabular}{cc}
   \includegraphics[width=0.7\hsize,angle=-90]{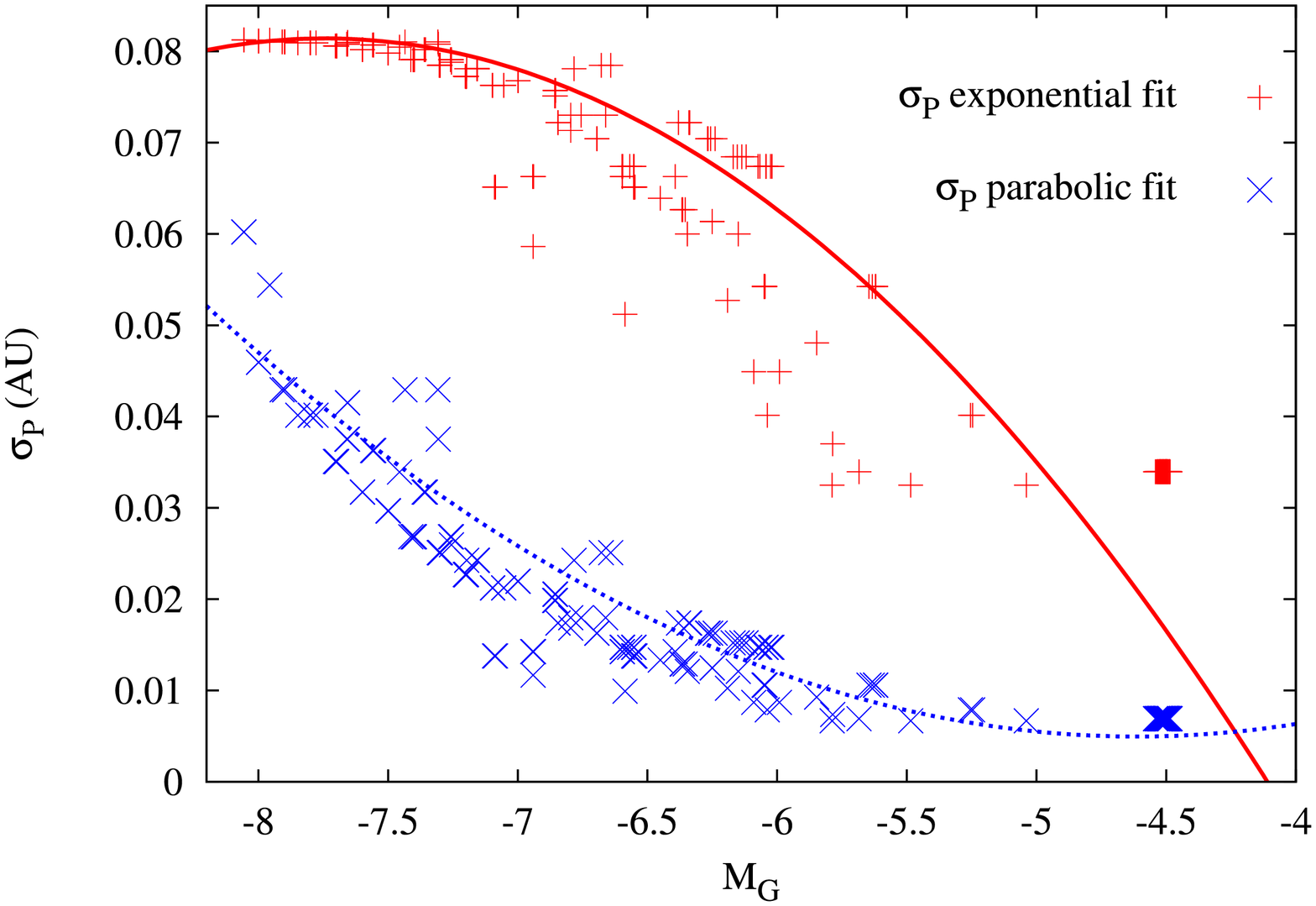}\\
   \includegraphics[width=0.7\hsize,angle=-90]{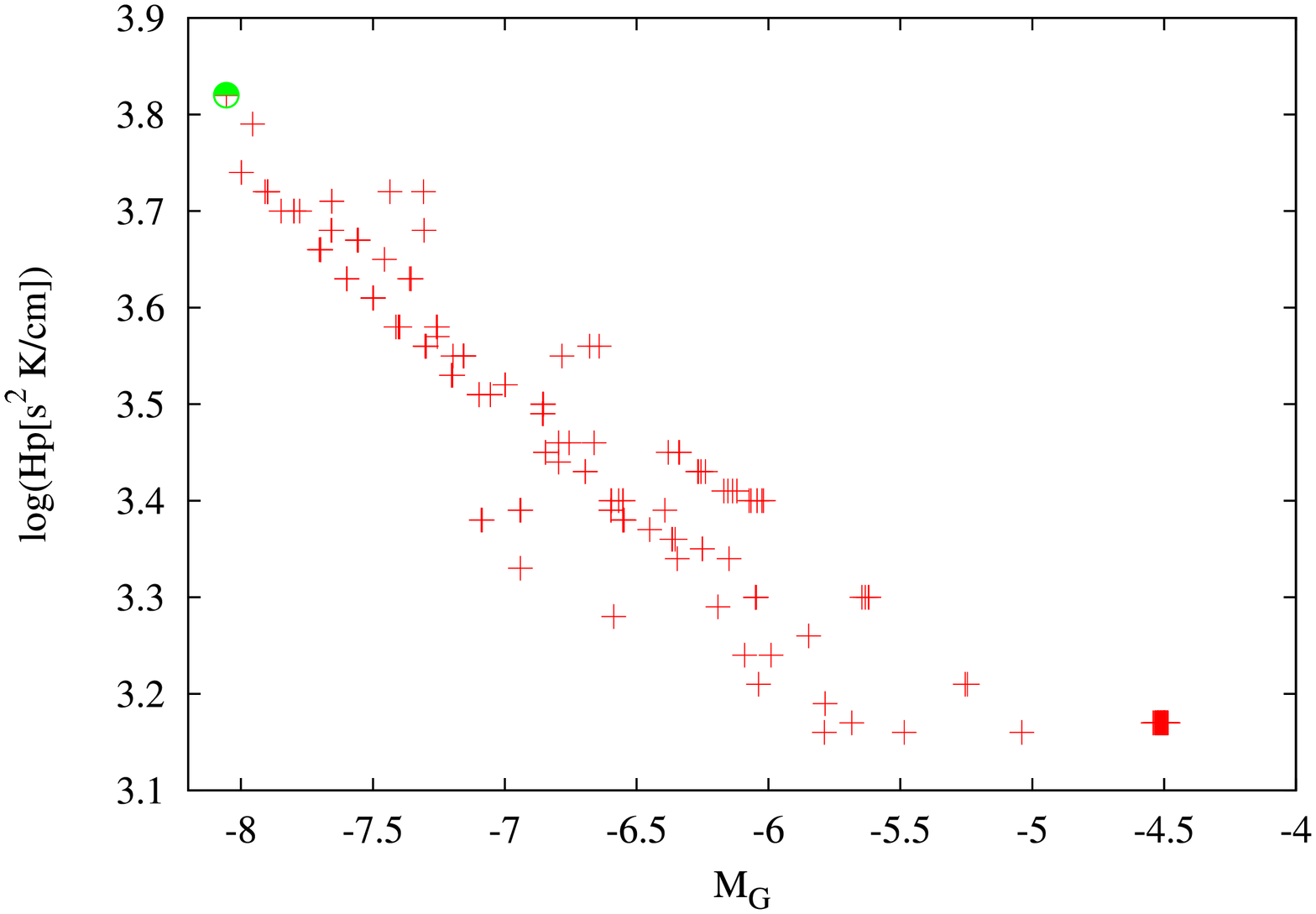}
\end{tabular}
      \caption{
Upper panel: the relation between $\sigma_{P_\theta}$  and $M_G$ for supergiants
and bright giants, assuming either an exponential (red plusses) or a parabola (blue crosses) to connect the box-in-a-star with the star-in-a-box models (Fig.~\ref{fig_gaia2bis}). The equation of the dashed green line (through the red plusses) is
$\sigma_{P_\theta}=-0.29-0.10\;M_G-0.0062\;M_G^2$, whereas the equation of the magenta dotted line (through the blue crosses) is 
$\sigma_{P_\theta}=0.083+0.034\;M_G+0.0037\;M_G^2$. 
Lower panel:  Same as the upper panel, for the relation between the pressure scale height $Hp$ and the absolute magnitude in the Gaia $G$ band. Only stars with a detectable photocentric motion (for the exponential fit: red crosses; for the parabolic fit: green dot) have been plotted.
}
         \label{Fig:MG-sigP}
   \end{figure}

The number of stars with photocentric motions detectable by Gaia as having bad fits (i.e., large goodness-of-fit $F2$ values) 
may now be estimated  as follows. The Besan\c con Galaxy model \citep{2004A&A...416..157R} has been used to generate a sample of  bright giants and  supergiants ($M_V < 0$) in the region $0\leq l \leq 180$ and $-20 \leq b \leq 20$ of our Galaxy (where $l$ and $b$ are the galactic coordinates).  
The reddening has been added separately using the extinction model from \citet{2003A&A...409..205D}.
For each one of the  the 361\ts069 stars in that sample, we assign the corresponding expected standard deviation of the 
photocenter displacement $\sigma_{P_\theta}$ taken from the
exponential or parabolic laws of Fig.~\ref{fig_gaia2} (each of these two possibilities being tested separately), with $H_P$ estimated from
Eq.~(\ref{Eq:HP}).\footnote{We note in passing that, in the Besan\c{c}on  sample, there is no star matching Betelgeuse parameters if one adopts $\log g =  -0.3$ for its surface gravity, yielding $\log H_p = 3.85$. If on the other hand, one adopts $\log g =  0.0$, we get $\log H_p=3.55$ and Betelgeuse is then matched by stars from the Besan\c{c}on  sample. This can be seen from the lower panel of Fig.~\ref{Fig:MG-sigP}, since Betelgeuse has
    $M_G=-6.4$, when adopting $M_{\rm bol}=-7.5$ from the apparent bolometric flux $111.67\times10^{-13}$~W~cm$^{-2}$ \citep{2004A&A...418..675P} and the parallax 6.56~mas \citep{vanLeeuwen-2007:a},  $V-G=0.98$ from $V-I=2.32$ \citep{ESA-1997} and Eq.~\ref{Eq:V-G}, $BC_V=-2.05$ from the apparent bolometric flux and $V = 0.42$ \citep{1966CoLPL...4...99J}.}

We then compute the number of stars which fulfill the condition expressed by Eq.~(\ref{Eq:M-d2-Gaia}), and having at the same time $G > 5.6$ in order 
not to saturate Gaia CCD detectors.  
The conversion between $V$ and $G$ magnitudes  has been done from the color equation 
(adopted from the {\it Gaia Science Performance} document\footnote{http://www.rssd.esa.int/index.php?project=GAIA 
\&page=Science\_Performance}$^,$\footnote{http://www.rssd.esa.int/SYS/docs/ll\_transfers/ project=PUBDB\&id=448635.pdf}):
\begin{eqnarray}
\label{Eq:V-G}
G &=& V - 0.0107 - 0.0879\; (V-I) - 0.1630\; (V-I)^2 \nonumber\\
   &   &+ 0.0086\; (V-I)^3.
\end{eqnarray}

With the exponential law, we found 215 supergiants (among the 361\ts069 of the full sample, representing half the galactic plane) fulfilling these two conditions. They are displayed in Fig.~\ref{fig_gaia1} in the $d - M_G$ plane, and are basically confined to a crescent delineated by the conditions $G > 5.6$ (corresponding to the two lines with an upward concavity, labelled $G > 5.6$; the two lines correspond to two values of the extinction in the $G$ band: $A_G = 0$ and 1) and 
$d  \le 55.5 \; \sigma_{P_\theta}(M_G)$ (Eq.~(\ref{Eq:M-d2-Gaia}), corresponding to the green dashed line with a downward concavity). The latter line is based on a fiducial relationship between $\sigma_{P_\theta}$  and $M_G$, as shown on Fig.~\ref{Fig:MG-sigP}. Some 
supergiants nevertheless fall outside the crescent defined above, simply because of the scatter affecting the $\sigma_{P_\theta} - M_G$ relationship (Fig.~\ref{Fig:MG-sigP}). Obviously, all the supergiants of interest are bright in the $G$ band, in the range 5.6 to about 8 and will thus be easily identifiable during Gaia data processing.

With the parabolic law, only one supergiant matches the conditions: it is the brightest supergiant located in the upper left corner of Fig.~\ref{Fig:MG-sigP} (green point in the lower panel; note that, in Fig.~\ref{fig_gaia1}, this star is not located below the parabolic threshold line as expected, because that line is based on a mean $\sigma_P - M_G$ relation -- see Fig.~\ref{Fig:MG-sigP} --, and that supergiant happens to have a $\sigma_P$ value much above average, as seen on Fig.~\ref{Fig:MG-sigP}). 
Thus, Fig.~\ref{fig_gaia1} suggests that the 'parabolic' 
link between box-in-a-star and star-in-a-box models of Fig.~\ref{fig_gaia2} and \ref{fig_gaia2bis} is a limiting case: for photocentric motions to be detected by Gaia, the $\sigma_{P_\theta}$ vs $M_G$ relation has to lie above this limiting case (depicted as the green solid line in Fig.~\ref{fig_gaia2bis}).

In Fig.~\ref{fig_gaia1}, there is a cluster of stars at $M_G = -4.5$ (corresponding to  $\log T_{\rm eff} \sim 3.5$ and $\log g \sim 0.4$) which corresponds to bright giants or asymptotic giant branch (AGB) stars. They are also clearly seen in Fig.~\ref{Fig:MG-sigP} as the cluster at  $\sigma_{P_\theta} = 0.035$~AU (with the exponential law) or 0.01~AU (with the parabolic law).
Since these stars belong to a population different from supergiants (with masses of the order 1~M$_\odot$), they are not necessarily confined to the galactic plane as supergiants are. Hence another sample, now covering a quarter of the sky ($0^\circ \le l \le 180^\circ$, $b \ge 0^\circ$), has been generated from the Besan\c{c}on model and contains 702211 giants and bright giants. In this sample, 938 stars satisfy the condition of detection of the photocentric motion with the exponential law, and none with the parabolic law. The  relation $M_G - \sigma_{P_\theta}$ thus appears as an essential ingredient, but unfortunately quite uncertain still, especially for those among the bright giants which are pulsating as long-period variables. The pulsation makes the modelling especially difficult
\citep[see for instance][for an application of 3D AGB models to the star VX~Sgr]{Freytag2008A&A...483..571F,Chiavassa_VLTI}. Nevertheless, numerous observations have revealed their surface brightness asymmetries \citep[e.g.,][and references therein]{2006ApJ...652..650R}.



\begin{table*}
\caption{Photocenter motion from 3D simulations. $H_{\rm{p}}$ is given with two different dimensions.
}             
\label{tab3d}      
\renewcommand{\footnoterule}{} 
\begin{tabular}{c c c c c r r r r r r rrr r}        
\hline\hline                 
Model & Configuration& $\log H_{\rm{p}}$ & $\log \mathcal{H}_{\mathrm{p}} $ &$\log(\sigma_{P_\theta})$ & $T_{\rm eff}$ & $\log g$ & $R$ & $M_{\rm bol}$ \\
 &  & (s$^2$K/cm) & ($10^7$ cm) & (AU) & (K) & & (R$_\odot$) \\
\hline
White dwarf$^a$ & box-in-a-star & -3.92 & -3.15 & -10.28 & 12000 & 8.00 & $1.28\;10^{-2}$ & 11.03\\
Sun$^{a}$ & box-in-a-star &-0.68          & 0.12 & -7.12 & 5780 & 4.44 & 1 & 4.74 \\
Procyon A$^{a}$ & box-in-a-star & -0.19    & 0.61    &-6.37 & 6540 & 4.00 & 2.10 & 2.59 & \\
$\xi$ Hydrae$^{a}$ & box-in-a-star &0.75    &   1.55     &-5.45 & 4880 & 2.94 & 10.55 & 0.36\\
Cepheid$^{a}$&box-in-a-star&1.66  & 2.46        &-4.25 & 4560 & 2.00 & 30.17 & -1.63\\
Red giant$^{a}$&box-in-a-star&2.57    &  3.36      &-3.02 & 3680 & 1.00 & 95.25 & -3.19 \\
RSG$^b$&star-in-a-box&  3.88&  4.68  &-1.10 & 3490 & -0.34 & 832 & -7.66 \\
\hline\hline                          
\end{tabular}

$^a$ \cite{2005ESASP.560..979S}\\
$^b$ This work, Sect.~\ref{sect:photocenter}\\
\end{table*}

\begin{table*}
\caption{\label{Tab:observP}
References used to compute the photocentric shifts from  interferometric data. 
Stellar parameters are from \cite{2005ApJ...628..973L} and \cite{2008AJ....135.1430H} for $\alpha$
Ori and $\alpha$ Her (set 2),  
\cite{1994A&A...285..915E} for $\alpha$ Her (set 1)
and \cite{1995AJ....110.2361F} for $o$ Cet. Parallaxes are  from \cite{vanLeeuwen-2007}. Only observations in the optical range have been kept.}
\begin{tabular}{rrrrrrrrrp{6.9cm}}
\hline
\hline
Name & $\log g$ & $T_{\rm eff}$ & $\log H_{\rm{p}}$ & $\varpi$ &\multicolumn{2}{c}{$P$}  & $\lambda$ &Date & References \\
\cline{6-7} 
     &          &  (K)          & log (s$^2$ K/cm) & (mas)   & (mas)& (AU)             & (nm)      &     &            \\  
\hline   
$\alpha$ Ori & $-0.3$ & 3650 & 3.86 &6.56&&&&&  \cite{2008AJ....135.1430H} (set 1)\\
                   &  0.0     & 3650 & 3.56 & 6.56 &&&&&   \cite{2005ApJ...628..973L} (set 2)\\
&&&&&1.216&			0.185& 700 & 02/1989 &   \cite{1990MNRAS.245P...7B},
\cite{1992MNRAS.257..369W,1997MNRAS.291..819W}, \cite{Tuthill-1997},
\cite{2000MNRAS.315..635Y}, \cite{2007ApJ...670L..21T},
\cite{Haubois2009}   \\
&&&&&1.637&			0.249& 710 & 01/1991& \\
&&&&&0.369&		    0.056& 700 & 01/1992 & \\
&&&&&0.694&			0.106& 700 & 01/1993 & \\
&&&&&0.550&			0.084& 700 & 09/1993 & \\
&&&&&0.427&			0.065& 700 & 12/1993 & \\
&&&&&0.144&			0.022& 700 & 11/1994 & \\
&&&&&0.395&			0.060& 700 & 12/1994 & \\
&&&&&0.302&			0.046& 700 & 12/1994 & \\
&&&&&0.142&			0.021& 700 & 01/1995 & \\
&&&&&0.025&			0.004& 700 & 01/1995 & \\
&&&&&0.009&			0.001& 700 & 11/1997 & \\
\cline{6-7}
 &&&&&$\langle P\rangle$& 0.075& && \\
 &&&&&$\langle P^2\rangle$ & 0.011 && & \\
&&&&&$\sigma_P$ & 0.075& &&  \\
\medskip\\
$\alpha$ Her& 0.76& 3400 & 2.77 & 9.07 &&& & &El Eid (1994)  (set 1)  \\
& 0.0 & 3450 & 3.53 & 9.07 &&& & & Levesque et~al. (2005) (set 2)  \\
&&&&&0.340&	0.037& 710 & 07/1992 & \cite{Tuthill-1997} \\
&&&&&0.765&	0.084& 710 & 06/1993 & \\
\cline{6-7}
& &&&&$\langle P\rangle$& 0.060 & & &\\
& &&&&$\langle P^2\rangle$&  0.004 & & & \\
&&&&&$\sigma_P$&  0.033& &  \\
\medskip\\
$o$ Cet & -0.6 & 2900 & 4.06 & 10.91 &&	& &  &\cite{1999MNRAS.306..353T}\\
&&&&&1.202&     0.110&710 & 07/1992& \\
&&&&&0.850&	0.078&700 & 01/1993& \\
&&&&&0.990&	0.091&710 & 09/1993& \\
&&&&&1.950&	0.179&710 & 12/1993& \\
\cline{6-7}
& &&&&$\langle P\rangle$ & 0.114\\
& &&&&$\langle P^2\rangle$ & 0.015\\
&&&&&$\sigma_P$&  0.045\\
\medskip\\
\hline
\end{tabular}
\end{table*}

\subsection{Impact on the parallaxes}\label{Sect:Gaia-parallax}

\begin{figure}
  \includegraphics[width=0.7\hsize,angle=-90]{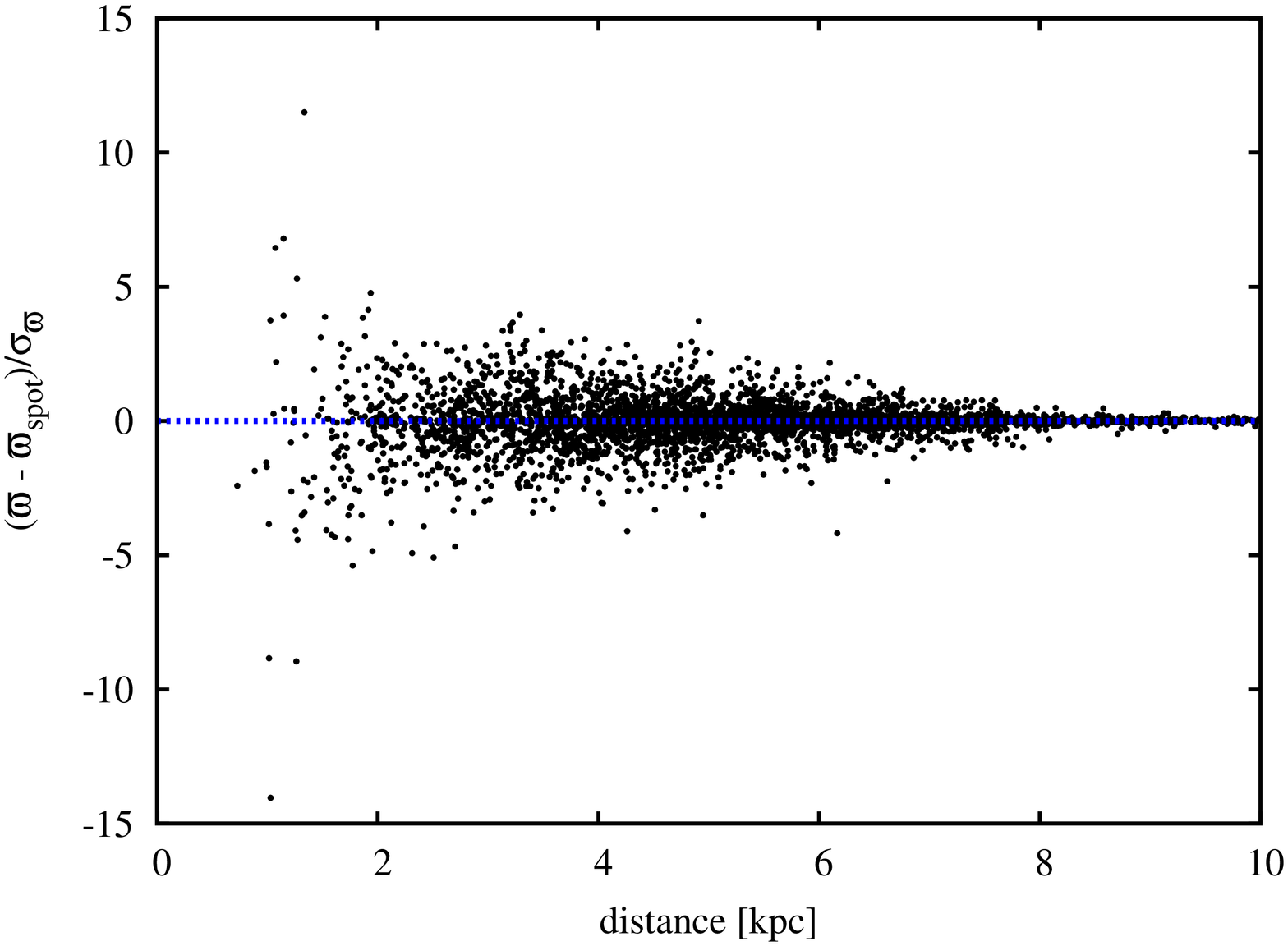} 
\caption{\label{Fig:parallax}
Comparison of parallaxes for supergiant stars with and without surface brightness asymmetries (spots), normalised to  $\sigma_\varpi$. The stars falling on the horizontal line with ordinate 0 are
very reddened stars, are consequently quite faint, and therefore have large errors on their astrometric measurements and thus on their parallax.
}
\end{figure}

To evaluate the impact of the photocentric shift on the parallax, we proceeded as follows. 
The sampling times, scanning angles, along-scan measurements and their errors were obtained from 
the Gaia Object Generator v7.0 \citep[GOG\footnote{http://gaia-gog.cnes.fr};][]{2010hsa5.conf..415I}
for the supergiant stars from the sample generated using the Besan\c con model described in the previous section.
A photocentric motion deduced from the photocentre position computed from the snapshots of the red supergiant model (see Fig.~\ref{fig3}) 
was added on the along scan measurements (the photocentric shift was converted from linear to angular shifts, according to the 
known stellar distance). The red supergiant model gives a single photocentre position sequence. Yet the sequence for every star should be different. Therefore the sequence was rotated for every star by a random angle before being added to the along scan measurements. The astrometric parameters were then retrieved by solving the least-squares 
equation (Eq.~(\ref{Eq:xi1})), separately with and without surface brightness asymmetries. The resulting parallaxes 
are compared in Fig.~\ref{Fig:parallax}. 

Fig.~\ref{Fig:histo} presents the histogram of the quantity $(\varpi-\varpi_{\rm spot})/\sigma_{\varpi}$ for three different ranges of distances. It is clearly seen that the distribution, quite peaked at zero for distant stars, becomes wider  for  nearer stars, meaning that the ratio of the error on the parallax to its formal error increases with decreasing distance. Similarly, the fits of the astrometric data are worse for stars closer by, and this effect is clearly seen on  Fig.~\ref{Fig:GoF}, displaying the relation between the goodness-of-fit parameter $F2$ and the distance. The run of $F2$ with distance is consistent with that predicted by Eq.~(\ref{Eq:DeltaF2}),  for $\nu= 70$, $\sigma_{P_\theta} = 0.1$~AU and $\sigma_\eta = 0.03$~mas.

\begin{figure}
  \includegraphics[width=0.7\hsize,angle=-90]{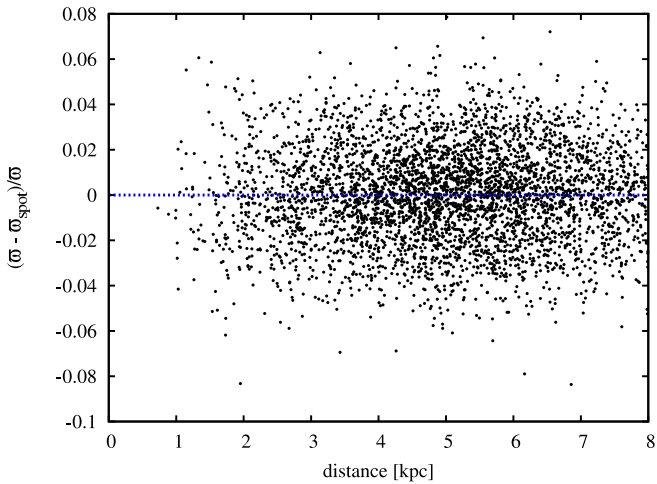} 
\caption{\label{Fig:parallaxrel}
Same as Fig.~\ref{Fig:parallax}, but normalized by the parallax. Note how the relative parallax error is almost independent of the distance. 
}
\end{figure}

\begin{figure}
  \includegraphics[width=0.9\hsize]{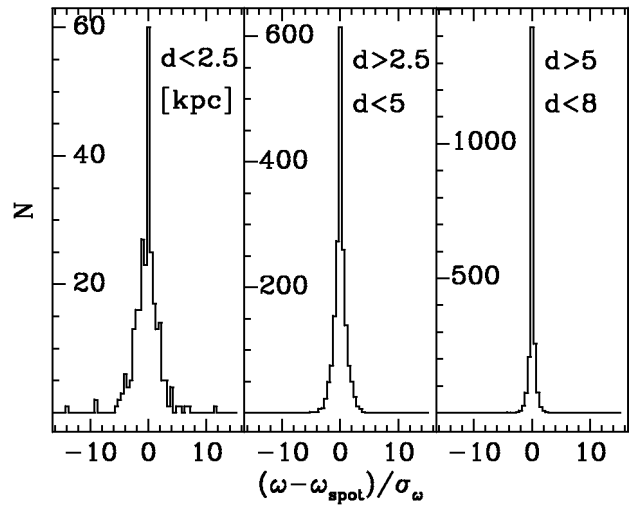} 
\caption{\label{Fig:histo}
Histograms of the relative error on the parallax of supergiant stars, for different ranges of distances.
}
\end{figure}

\begin{figure}
  \includegraphics[width=1.0\hsize]{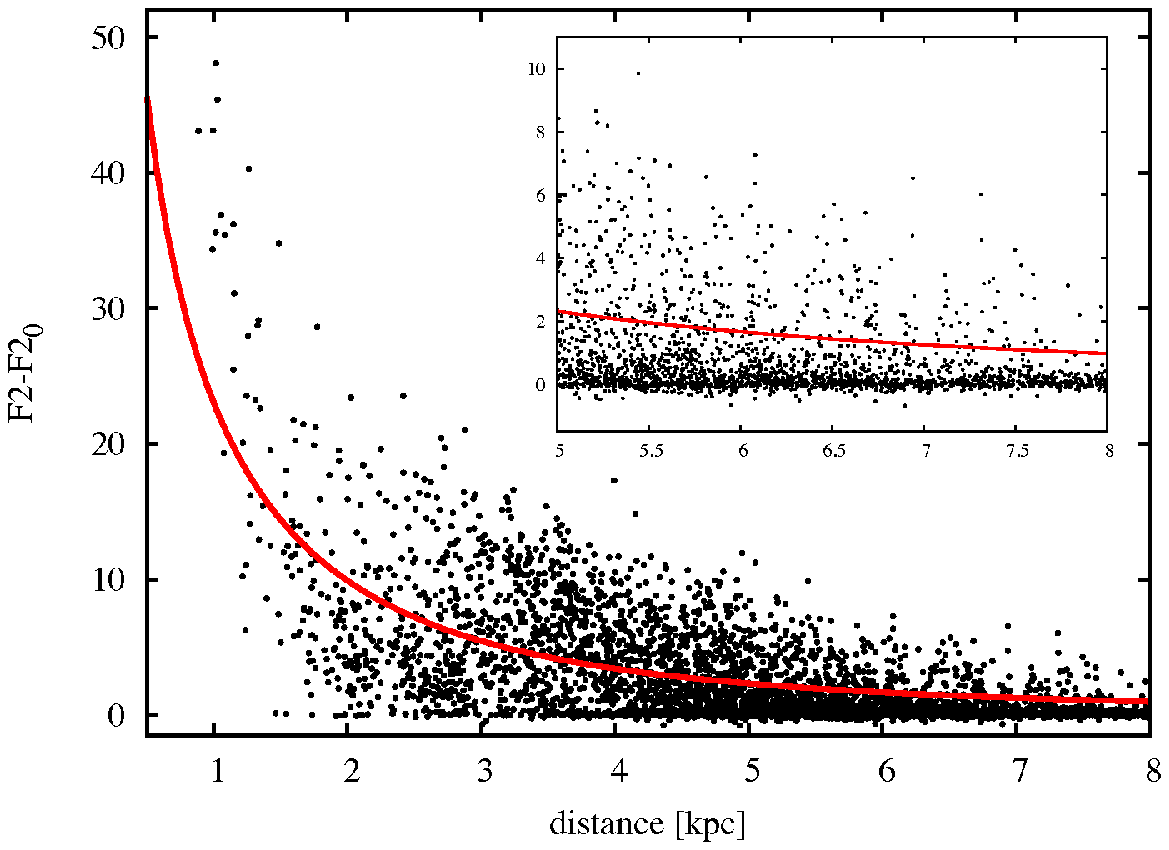} 
\caption{\label{Fig:GoF}
Same as Fig.~\ref{Fig:parallax} for the goodness-of-fit $F2$. The solid line corresponds to the prediction from 
Eq.~(\ref{Eq:DeltaF2}), with $\nu= 70$, $\sigma_{P_\theta} = 0.1$~AU and $\sigma_\eta = 0.03$~mas. Part of the large scatter at intermediate distances (2 to 4 kpc)
is due to different CCD gating sequence \citep[see][]{2005ESASP.576...35D,2010IAUS..261..296L}. 
}
\end{figure}

Coming back to Fig.~\ref{Fig:parallaxrel}, it is remarkable that the relative error on the parallax, namely $(\varpi-\varpi_{\rm spot})/\varpi$ is almost independent of the distance and amounts to a few percents. This is in fact easy to understand, if one assumes that the difference $\varpi-\varpi_{\rm spot}$ must somehow be proportional to the amplitude of the excursion of the photocenter on the sky, which must in turn be related to $\theta$, the angular radius of the star on the sky; therefore, $(\varpi-\varpi_{\rm spot}) /\varpi= \alpha \theta /\varpi = \alpha R$, where $\alpha$ is the proportionality constant and $R$ is the linear radius of the star (expressed in AU). Thus we conclude that the relative error on the parallax 
is independent of the distance, and is simply related to the excursion of the photocenter expressed in AU.


These simulations for a sample of Betelgeuse-like supergiants  thus allow us to confirm the results obtained in Sect.~\ref{Sect:Gaia-frequency} (and Fig.~\ref{fig_gaia1}), in particular the fact that the impact on the goodness-of-fit
remains noticeable up to about 5 or 6 kpc (Fig.~\ref{Fig:GoF}). 

\section{Conclusion}

We have provided astrometric and photometric predictions from 3D simulations of RSGs to evaluate the impact of 
the surface brightness variations on the astrometric parameters of these stars to be derived by Gaia.

We found that the global-scale convective pattern of RSGs cause strong
variability in the position of the photocenter, $P$. From a 3D simulation of a Betelgeuse-like supergiant, $\langle P\rangle=0.132\pm0.065$~AU
(i.e., more than 3$\%$ of the stellar radius) showing excursions from
0.005 to 0.3 AU over the 5 years of simulation. In addition, the spectra show large fluctuations in the red and blue Gaia bands of up to 0.28 mag in the blue and 0.15 mag in the red. The Gaia color index (blue - red) also fluctuates strongly with respect to time. Therefore, the uncertainties on [Fe/H], $T_{\rm eff}$  and $\log g$ should be revised upwards for RSGs due to
their convective motions. We have furthermore provided predictions for interferometric observables in the Gaia filters that can be tested against observations with interferometers such as VEGA at CHARA. 

Then we studied the impact of the photocentric noise on the astrometric parameters. For this purpose, we considered the standard deviation of the photocenter displacement predicted by the RHD simulation, sampled as Gaia will do (both timewise and directionwise). We called this quantity $\sigma_{P_{\theta}}$, where $\theta$ is the position angle of the scanning direction on the sky, and we found  $\sigma_{P_{\theta}}=0.08$~AU for Betelgeuse-like supergiants. This photocentric noise can be combined with $\sigma_\eta=30$ $\mu$as (the error on the along-scan position $\eta$) for Gaia to determine the maximum distance ($d<4.4$~kpc)  up to which a photocentric motion with $\sigma_{P_{\theta}}=0.08$~AU will generate an astrometric noise of the order of the astrometric error on one measurement (more precisely 0.6 times that error, yielding an increase of the $F2$ goodness-of-fit parameter by 2 units). The value $\sigma_{P_{\theta}}=0.08$~AU could even be somewhat underestimated, as we guessed from the comparison of the along-scan Hipparcos residuals for Betelgeuse with the  RHD predictions. We
concluded that the predicted photocentric noise does account for a
substantial part of the Hipparcos 'cosmic noise' for Betelgeuse and Antares, but not for all of it. This may be due to the fact that the temperature stratification in the RHD models is not completely correct due to the grey approximation used for the radiative transfer. The resulting intensity maps have higher contrast than the observations, as shown in Paper~II, and the photocenter position can thus also be affected. New simulations with
wavelength resolution (i.e., non-grey opacities) are in progress and
they will be tested against these observations.

We  estimated  how many RSGs might have have an abnormally large goodness-of-fit parameter $F2$. We found that  the photocentric noise should be detected by Gaia for a number of bright giants and supergiants varying between 2
and about 4190 (215 supergiants in each half of the celestial sphere and 940
 bright giants in each quarter of the sphere; see Sect.~\ref{Sect:Gaia-frequency}),  depending 
 upon the run of
$\sigma_{P_{\theta}}$ with the atmospheric pressure scale height $H_P$, and to a lesser extent, depending on galactic extinction. 
The theoretical predictions of 3D simulations presented in this
work will be tested against the multi-epoch interferometric
observations of a sample of giants and supergiants (Sacuto et al. in
preparation), with the hope to better constrain this $\sigma_{P_{\theta}} - H_P$ relation. In a forthcoming paper (Pasquato et al., in preparation), we will evaluate how the Gaia reduction pipeline behaves when facing the bright-giants and supergiants granulation. More specifically, we will show  that the distance to the star is the main driver fixing which one among all the possible solution types (single-star, acceleration, orbital, stochastic)  is actually delivered by the pipeline (the acceleration and orbital solutions being obviously spurious).

Finally, a very important conclusion is that the parallax for Betelgeuse-like supergiants may be affected by an error of  a few percents. For the closest supergiants ($d < 2.5$~kpc), this error may be up to 15 times the formal error $\sigma_{\varpi}$ (see Fig.~\ref{Fig:parallax}) resulting from the measurement errors and estimated from the covariance matrix. In a forthcoming paper (Pasquato et al., in preparation), we will moreover show that this error is sensitive to the time scale of the photocentric motion (which is in turn fixed by the granulation and the stellar rotation).

There is little hope to be able to correct the Gaia parallaxes of RSGs from this parallax error, without knowing the run of the photocentric shift for each considered  star. Nevertheless, it 
might be of interest  to monitor the photocentric deviations for a few well selected RSGs during the Gaia mission. Ideally, this would require imaging the stellar surface,  although monitoring of the phase closure 
on three different base lines may already provide valuable information on the size of the inhomogeneities present on the stellar surface (see Sacuto et al., in preparation).
The best suited targets 
for that purpose would be supergiants with $G$ magnitudes just above the Gaia saturation limit of 5.6, where the astrometric impact  is going to be maximum, and at the same time, still within reach of the interferometers.  The corresponding diameter will be on the order of 4 mas (derived from the radius 830\,$R_\odot$ for a Betelgeuse-like supergiant seen at a distance of 2~kpc if $G = 5.6$, $A_G = 1$, and $M_G = -6.6$). A search for G, K or M supergiants (of luminosity classes I, Ia, Iab or Ib) with $5.6 \le V \le 8$ in the SIMBAD database yielded only three stars (XX Per, HD 17306 and WY Gem) matching these criteria, the latter being a spectroscopic binary which will disturb the radius measurement and is thus unsuited for this purpose. 
It may therefore be necessary to select such targets from the Gaia data themselves, after the first year of the mission.

\begin{acknowledgements}
E.P. is supported by the ELSA (European Leadership in Space Astrometry) Research
Training Network of the FP6 Programme. S.S. acknowledges funding by the Austrian Science Fund FWF
under the project P19503-N13. We thank the CINES for providing some of the computational resources
 necessary for this work. We thank DPAC-CU2, and especially  X. Luri and Y. Isasi,  for help with the use of  GOG. A.C. thanks G. Jasniewicz for enlightening discussions. B.F. acknowledges financial support from
the {\sl Agence Nationale de la Recherche} (ANR),
the {\sl ``Programme National de Physique Stellaire''} (PNPS) of CNRS/INSU,
and the {\sl ``{\'E}cole Normale Sup{\'e}rieure''} (ENS) of Lyon, France,
and from the {\sl Istituto Nazionale di Astrofisica / Osservatorio Astronomico di Capodimonte}
(INAF/OAC) in Naples, Italy.
\end{acknowledgements}

\bibliographystyle{aa}
\bibliography{biblio.bib}

\begin{appendix} 
\section{Formal errors on the parameters of a least-squares minimisation}
We provide here a short demonstration of a well-known 
statistical result \citep[see e.g., ][]{Press-1992}, which may appear counter-intuitive in the present context, namely the fact that the presence of an extra-source of unmodelled noise will not change the formal errors on the parameters derived from a least-squares minimisation. 

Consider the case where the data points $(x_i, y_i) \;(i=1,...N) $ must be fitted by a general linear model
\begin{equation}
y(x) = \Sigma_{k=1}^M a_k\;X_k(x)
\end{equation}
where $X_k(x)\; (k=1,...M)$ are $M$ arbitrary (but known) functions of $x$, which may be wildly non linear. The merit function is defined as 
\begin{equation}
\chi^2 = \sum_{i=1}^{N} \left[\frac{y_i - \Sigma_{k=1}^M a_k\;X_k(x_i)}{\sigma_i}  \right]^2 ,
\end{equation} 
where $\sigma_i$ is the measurement error on $y_i$ presumed to be known. To simplify the notation, we define the {\it design matrix} {\bf A} (of size $N \times M$)  by 
\begin{equation}
A_{ij} = \frac{X_j(x_i)}{\sigma_i},
\end{equation} 
the vector $\mathbf{b}$ of (normalized) measured values, of length $N$:
\begin{equation}
b_i = \frac{y_i}{\sigma_i},
\end{equation}
and finally the vector $\mathbf{a}$ of length $M$ whose components are the parameters $a_k \; (k=1,...M)$ to be fitted.
The least-squares problem may thus be rephrased as
\medskip\\
\begin{center}
find $\mathbf{a}$ that minimizes $\chi^2 = \left| \mathbf{A} \cdot \mathbf{a} - \mathbf{b}\right|^2$, 
\end{center}
whose solution may be written
\begin{equation}
\mathbf{\left( A^T \cdot A \right) \cdot a = A^T \cdot b},
\end{equation}
with  $\mathbf{C \equiv \left(A^T \cdot A\right)^{-1}}$ being the variance-covariance matrix  describing the uncertainties\footnote{In fact, this statement 
only holds in the case where the errors $\sigma_i$ are normally distributed, which is supposed to be the case for the specific problem under consideration (namely, the Gaia along-scan measurement errors).} of the estimated parameters ${\mathbf a}$. The crucial point to note here is the fact that matrix ${\mathbf C}$ involves the measurement uncertainties $\sigma_i$ but not the measurements $y_i$ themselves. Therefore, changing $y_i$, in the presence of an unmodelled process (like photocentric motion) without changing the measurement uncertainties $\sigma_i$,  will not change the formal errors on the resulting parameters $\mathbf{a}$. But of course, $\chi^2$ along with the goodness-of-fit parameter $F2$ (see
Eq.~(\ref{Eq:F2})) will be larger in the presence of a
photocentric noise, as the scatter around the best astrometric solution will be larger
    than expected solely from the measurement errors.
    Therefore, it is $F2$ and its associated $\chi^2$, but not the formal parallax error,  which  bear the signature
of the presence of photometric noise.

\end{appendix}

\end{document}